\documentclass{aa}
\usepackage[dvips]{graphicx}
\usepackage{txfonts}
\usepackage{longtable}
\def\HII{H{\sc ii} }

\def\UC{UC~H{\sc ii}}

\def\kms{\mbox{km~s$^{-1}$}}

\def\juz{\mbox{1--0}}

\def\hi{Hi-GAL}
\def\g29{G29.96$-$0.02}

\begin{document}
\title{A \hi\ study of the high-mass star-forming region G29.96$-$0.02} 
\author{M.\ T.\ Beltr\'an\inst{1}, L.\ Olmi\inst{1, 2}, R.\ Cesaroni\inst{1}, 
 E.\ Schisano\inst{3}, D.\ Elia\inst{3},  S.\ Molinari\inst{3}, 
  A.\ M.\ Di Giorgio\inst{3}, J.\ M.\ Kirk\inst{4}, J.\ C.\ Mottram\inst{5}, 
 M.\ Pestalozzi\inst{3}, L.\ Testi\inst{1, 6}, \and M.\ A.\ Thompson\inst{7}}
\institute{
INAF-Osservatorio Astrofisico di Arcetri, Largo E.\ Fermi 5,
50125 Firenze, Italy
\and
University of Puerto Rico, R\'io Piedras Campus, Physics Dept., Box
23343, UPR station, San Juan, Puerto Rico, USA
\and
INAF-Istituto di Astrofisica e Planetologia Spaziali, 
via del Fosso del Cavaliere 100, 00133 Roma, Italy
\and
Jeremiah Horrocks
Institute, University of Central Lancashire, Preston PR1 2HE, UK
\and
Leiden Observatory, Leiden University, PO Box 9513, 2300 RA Leiden, The
Netherlands
\and
ESO, Karl Schwarzschild str.\ 2, 85748 Garching, Germany
\and
Centre for Astrophysics Research, STRI, University of
Hertfordshire, College Lane, Hatfield, AL10 9AB, UK
}
\offprints{M.\ T.\ Beltr\'an, \email{mbeltran@arcetri.astro.it}}
\date{Received date; accepted date}

\titlerunning{\hi\ sources in the G29.96$-$0.02 cloud}
\authorrunning{Beltr\'an et al.}

\abstract
{G29.96$-$0.02 is a high-mass star-forming cloud observed at 70, 160, 250, 350,
and 500~$\mu$m as part of the {\it Herschel} {\rm survey of the Galactic Plane (\hi) during 
the Science Demonstration Phase.}}
{\rm We wish to conduct a far-infrared study of the sources associated with this
star-forming region by estimating their physical properties and evolutionary
stage, and investigating the clump mass function, the star formation efficiency 
and rate in the cloud.}
{We have identified the \hi\ sources associated with the cloud, searched for possible counterparts at
centimeter and infrared wavelengths, fitted their 
spectral energy distribution and estimated their physical parameters.}
{A total of 198 sources have been detected in all 5 \hi\ bands, 117 of which
are associated with 24~$\mu$m emission and 87 of which are not associated with 24~$\mu$m
emission. We called the former sources 24~$\mu$m-bright and the latter ones
24~$\mu$m-dark. The [70--160] color of the 24~$\mu$m-dark sources is smaller than that
of the 24~$\mu$m-bright ones. The 24~$\mu$m-dark sources have 
lower $L_{\rm bol}$ and $L_{\rm bol}/M_{\rm env}$ than the 24~$\mu$m-bright ones for similar 
$M_{\rm env}$, which suggests that they are in an earlier evolutionary phase. The G29-SFR cloud 
is associated with 10 NVSS sources and with extended centimeter
continuum emission well correlated with the 70~$\mu$m emission. Most of the NVSS sources
appear to be early B or late O-type stars. The most massive and luminous \hi\ sources in the
cloud are located close to the G29-UC region, which suggests that there is a
privileged area for massive star formation towards the center of the G29-SFR
cloud. Almost all the \hi\ sources have masses well above the Jeans
mass but only 5\% have masses above the virial
mass, which indicates that most of the sources are stable against gravitational
collapse. The sources with $M_{\rm env}>M_{\rm virial}$ and that should be 
undergoing collapse and forming stars  are preferentially located 
at $\lesssim 4'$ of the G29-UC region, which is the most luminous source in the cloud. 
The overall SFE of the G29-SFR cloud ranges from 0.7 to 5\%, and the 
SFR  ranges from 0.001 to 0.008~$M_\odot$\,yr$^{-1}$, consistent with the values
estimated for Galactic \HII regions. The mass spectrum of the sources with masses above $300~M_\odot$, well above the
completeness limit, can be well-fitted with a power law of slope
$\alpha=2.15\pm0.30$, consistent with the values obtained for the whole 
$l=30\degr$, associated with high-mass star formation, and $l=59\degr$,
associated with low- to intermediate-mass star formation, \hi\ SDP fields.}
{}

\maketitle

\section{Introduction}

\begin{figure*}
\centerline{\includegraphics[angle=-90,width=17.5cm]{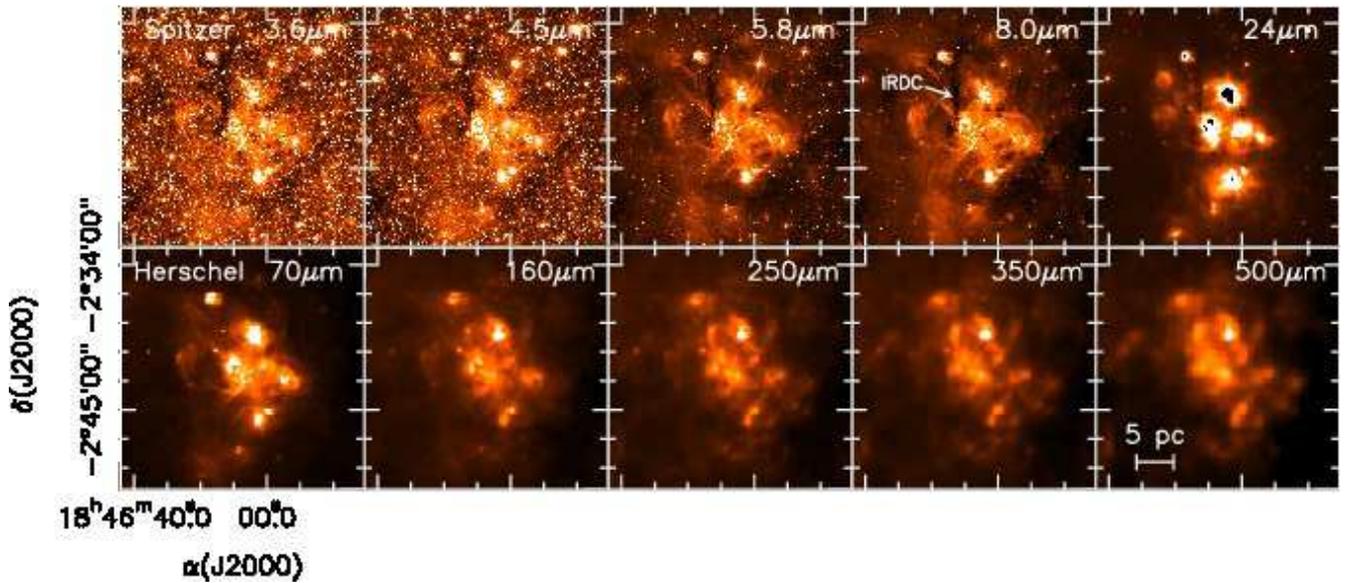}}
\caption{{\it Spitzer} 3.6, 4.5, 5.8, 8.0, and 24~$\mu$m and {\it Herschel} 70,
160, 250, 350, and 500~$\mu$m images in linear scale of the G29-SFR cloud. The white arrow in the
8.0~$\mu$m image points to the filamentary IRDC (Pillai et al.~\cite{pillai11}).}
\label{ir}
\end{figure*}

The G29.96$-$0.02 star-forming region (hereafter G29-SFR), located at a distance of
6.2~kpc (Russeil et al.~\cite{russeil11}), is a well-studied high-mass
star-forming cloud which falls in one of the two Science Demonstration
Phase (SDP) fields observed by the  ESA {\it Herschel} Space Observatory (Pilbratt et
al.~\cite{pilbratt10}) for the {\it Herschel} Infrared GALactic plane survey
(\hi: Molinari et al.~\cite{molinari10}). \hi\ is a {\it Herschel} key project
aimed at mapping the Galactic plane in five photometric bands (70, 160, 250,
350, and 500~$\mu$m). Figure~\ref{ir} shows the cloud as seen in different
wavelengths, from 3.6 to 500~$\mu$m, by {\it Spitzer} and {\it Herschel}. 

This cloud is dominated by IRAS~18434$-$0242, the brightest source from 24 to
500~$\mu$m (Fig.~\ref{ir}; Kirk et al.~\cite{kirk10}), and one of the brightest radio and infrared sources
in the Galaxy. This source is associated with a cometary \UC\ region (hereafter 
G29-UC: Cesaroni et al.~\cite{cesa94}; De Buizer et al.~\cite{debuizer02}) and
with a Hot Molecular Core (hereafter G29-HMC) located right in front of the
cometary arc (Wood \& Churchwell~\cite{wood89}; Cesaroni et al.~\cite{cesa94},
\cite{cesa98}). The G29-HMC core, which has been mapped in several tracers
(Cesaroni et al.~\cite{cesa98}; Pratap et al.~\cite{pratap99}; Maxia et
al.~\cite{maxia01}; Olmi et al.~\cite{olmi03}; Beuther et al.~\cite{beuther07};
Beltr\'an et al.~\cite{beltran11}), shows a velocity gradient approximately
along the east-west direction, which has been interpreted as rotation of a  huge
and massive toroid (4000~AU of radius and 88~$M_\odot$ at a distance of 6.2~kpc:
Beltr\'an et al.~\cite{beltran11}).

The G29-SFR cloud also contains a filament seen in absorption in the {\it
Spitzer} images (Fig.~\ref{ir}) and in emission in the SCUBA Massive Pre-/Proto-cluster core
Survey (SCAMPS: Thompson et al.~\cite{thompson05}) at about $2'$ east of the
G29-UC region (see {\it Spitzer}  image at 8~$\mu$m in Fig.~\ref{ir}). This
Infrared Dark Cloud (IRDC) has been extensively studied at high-angular
resolution in dust continuum emission and NH$_2$D by Pillai et
al.~(\cite{pillai11}), who have resolved, with an angular resolution better than
$5''$, the dust and line emission of the filament into multiple massive cores
with low temperatures, $<20$~K, and a high degree of deuteration. These findings
support the idea that this massive IRDC is in a very early stage of evolution,
and could be in a pre-cluster phase. Only the  brightest millimeter continuum core
shows signs of high-mass star-formation activity, as indicated by the point
source already visible at 24~$\mu$m that is driving a molecular outflow. That no
active star formation has been detected in other parts of this IRDC (Pillai et
al.~\cite{pillai11}) supports the idea of this extincted filament being in a
very early evolutionary phase.

As just seen, the G29-SFR cloud represents an ideal laboratory to study star
formation because young stellar objects in different evolutionary stages and
different masses are embedded in it. In this paper, we present a far-infrared
(FIR) study of this cloud using the \hi\ data in the 2 PACS and 3 SPIRE
photometric bands, centered at 70, 160, 250, 350, and 500~$\mu$m. Our goal is to
identify the FIR sources associated with this high-mass star-forming region and
estimate their physical properties (mass, temperature, luminosity, and density)
together with the Clump Mass Function (CMF) of the cloud. Combining the data
with ${\it Spitzer}$ and radio continuum observations, we will investigate the
evolutionary stage of the sources and their distribution in the cloud, and the
physical parameters of the associated \HII regions. Finally, we will derive the
star formation efficiency and star formation rate in this cloud. This work
complements the other wide-field studies carried out as part of the Hi-GAL SDP
(e.g.\ Bally et al.~\cite{bally10}; Battersby et al.~\cite{battersby11}; Olmi et
al.~\cite{olmi13}).

\section{Source selection}

\begin{figure*}
\centerline{\includegraphics[angle=-90,width=14.5cm]{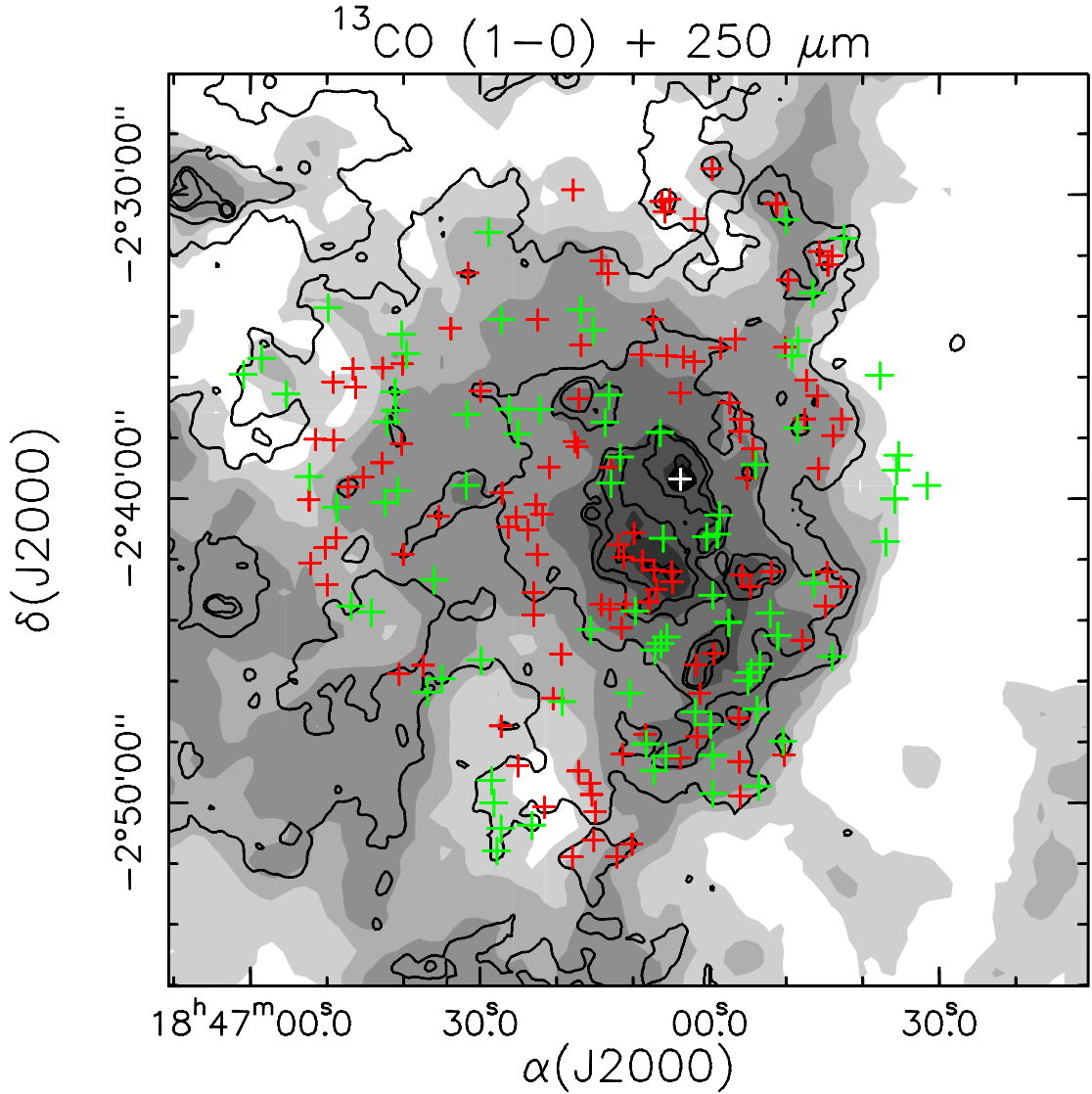}}
\caption{Hi-GAL 250~$\mu$m emission ({\it contours}) of the G29-SFR cloud 
overlaid on the 
$^{13}$CO~(\juz) emission ({\it grayscale}) from GRS averaged over the 95--105~\kms\ velocity
interval. Contour levels are 7, 9, 14, 18, 24, 36, 48, 72, and 96$\%$ of the 
peak of the 250~$\mu$m emission, 36311~MJy/sr. Grayscale levels are 10, 15, 20, 40, 60, 80, and
100$\%$ of the peak of the $^{13}$CO emission, 5~K. The crosses indicate the positions of
the \hi\ sources
associated with the cloud detected in the 5 \hi\ photometric bands. Red and
green crosses indicate, respectively, sources associated and non-associated 
with {\it Spitzer} 24~$\mu$m emission. The white
cross marks the position of the \hi\ source associated with the G29-UC region 
and G29-HMC core (see
$\S$~\ref{hmc}), which is also associated with 24~$\mu$m  emission. 
}
\label{co-sou}
\end{figure*}

The first step to identify the Hi-GAL sources associated with the G29-SFR cloud
is to define the limits of the molecular cloud. To study the distribution of the
gas in the region we have used the $^{13}$CO~(\juz) data of the Boston
University--Five College Radio Astronomy Observatory Galactic Ring Survey (GRS:
Jackson et al.~\cite{jackson06}). Towards the direction of the G29-UC region,
the $^{13}$CO~(\juz) emission shows relatively narrow components at $\sim$8, 49,
and 68~\kms, and a much broader component from $\sim$90 to 110~\kms. Taking into
account that the systemic velocity of high-density tracers, such as NH$_3$ or
CH$_3$CN, observed towards the G29-HMC core is $\sim$98--99~\kms\ (Cesaroni et
al.~\cite{cesa98}; Beltr\'an et al.~\cite{beltran11}), we selected the latter
broad velocity component to determine the distribution of the gas in the cloud.
The $^{13}$CO~(\juz) emission has been averaged over the 95--105~\kms\ velocity
interval and compared with the Hi-GAL 250~$\mu$m emission. As one can see in
Fig.~\ref{co-sou}, the gas and dust emission are very well correlated. The
G29-SFR cloud has been defined as the region contained approximately within the
contour at 10-15$\%$ of the $^{13}$CO peak emission (5~K) and that at 7$\%$ of
the 250~$\mu$m peak emission (36311~MJy/sr). Only the \hi\ sources falling
inside this region have been assigned to the G29-SFR cloud. 

\subsection{Source extraction} 

The source extraction and brightness estimation techniques applied to the \hi\
maps in this work are similar to the methods used during analysis of the BLAST05
(Chapin et al.~\cite{chapin08}) and BLAST06 data (Netterfield et 
al.~\cite{netterfield09}; Olmi et al.~\cite{olmi09}). However, important
modifications have been applied to adapt the technique to the SPIRE/PACS maps. 
The method used here defines in a consistent manner the region of emission of
the {\it same volume} of gas/dust at different wavelengths, thus differing from
the source grouping and band-merging procedures described by Molinari et
al.~(\cite{molinari11}) and Elia et al.~(\cite{elia10}). Candidate sources are
identified by finding peaks after a Mexican Hat Wavelet type convolution is
applied to all five SPIRE/PACS maps. Initial candidate lists from 70, 160 and
$250\,\mu$m  are then found and fluxes at all three bands extracted by fitting a
compact Gaussian profile to the source. Sources are not identified at 350 and
$500\,\mu$m  due to the greater source-source and source-background confusion
resulting from the lower resolution, and also because these two SPIRE wavebands
are in general more distant from the peak of the source Spectral Energy
Distribution (SED). Each temporary source list at 70, 160 and $250\,\mu$m is
then purged of overlapping sources and then all three lists are merged. After
selecting the sources based on their integrated flux and allowed angular
diameter, a final source catalog is generated. In the next stage, Gaussian
profiles are fitted again to all SPIRE/PACS maps, including the 350 and
$500\,\mu$m wavebands, using the size and location parameters determined at the
shorter wavelengths during the previous steps (the size of the Gaussian is
convolved to account for the differing beam sizes). Since the volume of emission
is basically defined using the 250~$\mu$m band, this method does not fully
exploit the higher angular resolution available at the shortest wavelengths. The
interested reader can find more details in Olmi et al.~(\cite{olmi13}).

The total number of \hi\ sources associated with the G29-SFR cloud is 198. The
position of the sources in equatorial and galactic coordinates, their fluxes in
the 5 photometric Hi-GAL bands, and their possible association with MIPSGAL
24~$\mu$m sources are given in Table~1.

\begin{figure*}
\centerline{\includegraphics[angle=-90,width=15cm]{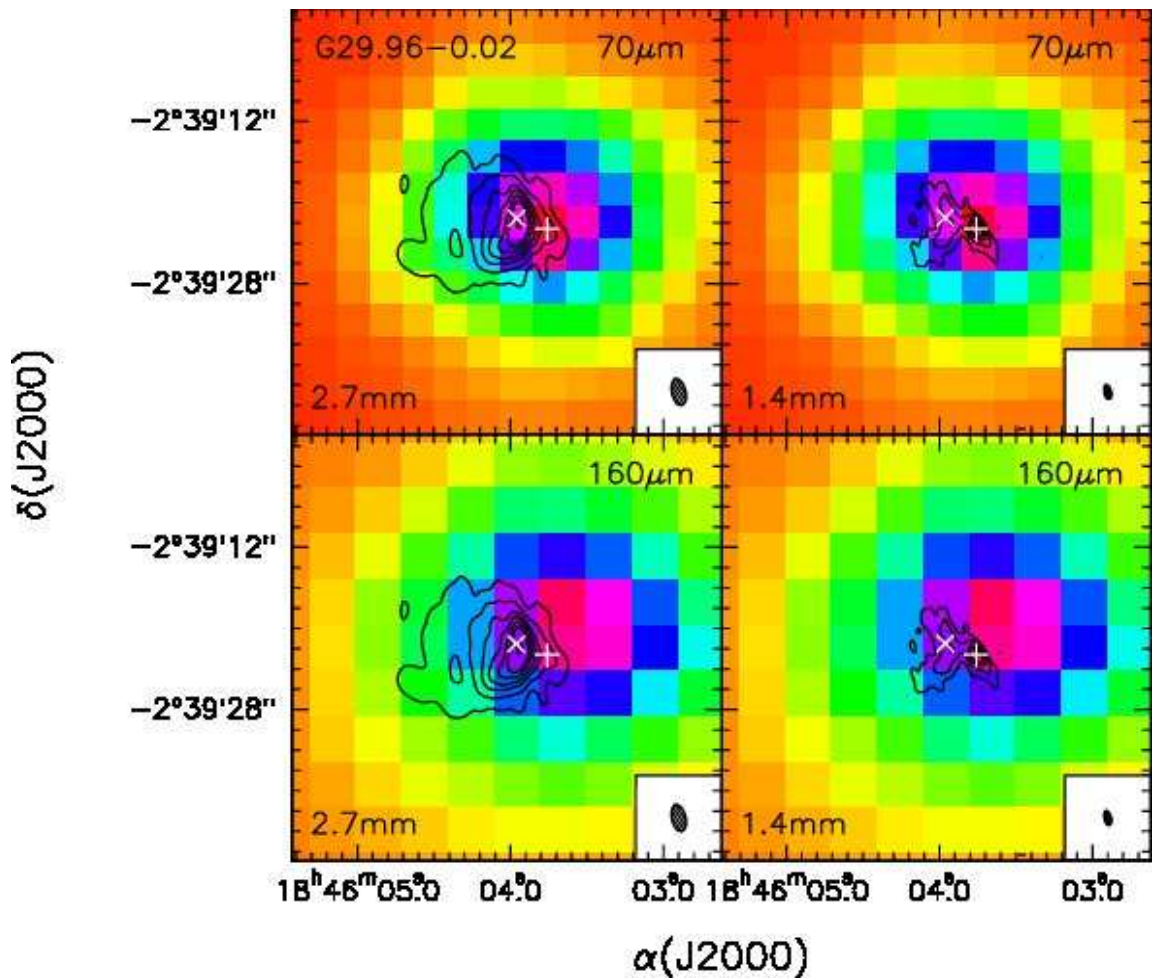}}
\caption{Overlay of the 2.7~mm ({\it left panels}) and  the 1.4~mm ({\it right panels})
continuum emission (contours) obtained with the PdBI
(Beltr\'an et al.~\cite{beltran11}) on the Hi-GAL 70~$\mu$m  
({\it upper panels}) and 160~$\mu$m ({\it lower panels}) emission (grayscale)
towards the position of the G29-UC region and the G29-HMC core. The white X marks
the position of the G29-UC region (Wood \& Churchwell~\cite{wood89}) and the 
white cross the
position of the 1.4~mm continuum emission peak associated with the G29-HMC core
(Beltr\'an et al.~\cite{beltran11}). The contour levels are 3, 9, 18, 27, 39,
51, and 75 times 3~mJy\,beam$^{-1}$ at 2.7~mm and 6.7~mJy\,beam$^{-1}$ at
1.4~mm. The PdBI synthesized beam
is shown in the lower righthand corner.}
\label{plateau}
\end{figure*}

\section{Analysis}

\subsection{Spectral Energy Distribution fitting}
\label{sedfit}

To estimate the dust temperature $T$, the mass $M_{\rm env}$, and the luminosity
$L_{\rm bol}$,
of the sources associated with the G29-SFR cloud, we fitted their observed
SED with a modified blackbody of the form
$B_\nu(T)(1-e^{-\tau_\nu})\Omega_S$, where $B_\nu(T)$ is the Planck function at
a frequency $\nu$ for a dust temperature $T$, $\tau_\nu$ is the dust optical
depth taken as $\tau_\nu\propto \nu^{\beta}$, where $\beta$ is the dust
emissivity index, and $\Omega_S$ is the source solid angle. The source size
$\theta$, which is not deconvolved, was estimated at 160~$\mu$m by the source extraction process (Olmi et
al.~\cite{olmi13}). The masses were calculated assuming
a dust mass absorption coefficient of 0.5 cm$^2$/g at 1.3~mm (Kramer et
al.~\cite{kramer03}) and a
gas-to-dust ratio of 100. To check whether the SED fitting
improved, we searched for counterparts of the \hi\ sources at shorter
wavelengths in the MIPSGAL 24~$\mu$m catalog (Shenoy et al.~\cite{shenoy12}). The method used to associate \hi\
and MIPSGAL sources, which was based on both a  positional and a color criteria,
is described by Olmi et al.~(\cite{olmi13}). For the remaining \hi\ sources or
those sources saturated at 24~$\mu$m, we searched for a counterpart in the
Wide-field Infrared Survey Explorer (WISE) catalog at 22~$\mu$m (Wright et
al.~\cite{wright10}). To associate a
WISE source to a \hi\ source, we arbitrarily chose the closest WISE source
located at $<12''$, the WISE angular resolution at 22~$\mu$m (Wright et
al.~\cite{wright10}). Finally, for the remaining \hi\ sources or those sources
saturated at 22~$\mu$m, we searched for a counterpart in the Midcourse Space
Experiment (MSX) catalog at 21~$\mu$m (Price et al.~\cite{price01}). In this
case,  we arbitrarily associated
the closest MSX source located at $<18\farcs3$, the MSX angular resolution at
21~$\mu$m (Price et al.~\cite{price01}). We found 103 MIPSGAL sources not
saturated at 24~$\mu$m,  11 WISE sources not saturated at 22~$\mu$m, and 6 MSX
sources associated with the \hi\ ones. The SED fitting was
performed using the 5 \hi\ bands for 157 sources. For these 157 sources  having a counterpart at shorter
wavelengths, including the additional point in the SED did not improve the
fitting. For 13 sources, only the 160, 250, 350, and 500~$\mu$m \hi\ bands were
used. For these sources, the flux at 160~$\mu$m, $S_{160 \mu m}$, was $\leq
S_{250 \mu m}$ and including the $S_{70 \mu m}$ in the SED clearly worsen the
fit. This indicates that the 70~$\mu$m emission is likely tracing a different
source component, more associated with the central stellar object, than that
traced by the emission at 160 to 500~$\mu$m, more associated with the extended
envelope surrounding the central source. The  5 \hi\ bands plus the
21~$\mu$m band of MSX were used in the SED fitting for 4 sources. In these 
cases, $S_{70 \mu m} > S_{160 \mu m}$ and including the flux at 21~$\mu$m, which is smaller than
that at 70~$\mu$m, clearly improved the fitting. For 6 sources, the 5 \hi\ bands
plus the 22~$\mu$m band of the WISE were used in the fitting. For these sources,
$S_{70 \mu m} > S_{160 \mu m}$ and $S_{70 \mu m} >S_{22 \mu m}$, and again,
including the flux at a shorter wavelength improved the fitting. Finally, for 18
sources, the 5 \hi\ bands plus the 24~$\mu$m MIPSGAL band were used.  For these
sources, $S_{70 \mu m} > S_{160 \mu m}$ and $S_{70 \mu m} >S_{24 \mu m}$, and as
in the previous cases, including the flux at a shorter wavelength improved the
fitting.  The MSX flux at 21~$\mu$m, the WISE flux at 22~$\mu$m, and the MIPSGAL
flux at 24~$\mu$m used for the SED fitting of these 28 sources is given in
Table~2. Table~3 shows the values of $\theta$, obtained from the source extraction
process,  of $\beta$, $T$, $M_{\rm env}$, and $L_{\rm bol}$, obtained from the SED fitting, and of
the surface density $\Sigma$, for the 198 sources. The surface density was
calculated following the expression $\Sigma=M_{\rm env}/(\pi\times R^2)$, where the 
radius of the sources $R$ was obtained from their sizes, $\theta$, and following the expression $R=\theta/2\times d$, where $d$ is the distance to the G29-SFR cloud. 

\addtocounter{table}{+3}
\begin{table*}
\caption[] {Mean (median) physical parameters of the \hi\ sources} 
\label{taverage}
\begin{tabular}{lccc}
\hline
&\multicolumn{1}{c}{All sources} &
\multicolumn{1}{c}{24$\mu$m-dark} &
\multicolumn{1}{c}{24$\mu$m-bright} 
\\
\hline
 Radius (pc) &  0.36 (0.36)  & 0.34 (0.35) & 0.37 (0.38) \\
 Mass ($M_\odot$) & 379 (115) & 435 (172) & 340 (86)  \\
 Surface density (g\,cm$^{-2}$) & 0.24 (0.06) & 0.27 (0.1) & 0.22 (0.04) \\ 
 Temperature (K) &  29 (25) & 22 (22)& 33 (30) \\
 Luminosity ($L_\odot$) &  $6.2\times10^3$ (470) & 706 (247) & $1\times10^4$ (713) \\
 Luminosity-to-mass ratio ($L_\odot/M_\odot$) & 23 (5) & 6 (2) & 34 (10) \\
\hline
\end{tabular}
\end{table*}

\subsection{The \hi\ source associated with the G29-UC region and G29-HMC core}
\label{hmc}

Figure~\ref{plateau} shows an overlay of the continuum emission at 2.7 and
1.4~mm obtained with the IRAM-Plateau de Bure interferometer (PdBI) (Beltr\'an
et al.~\cite{beltran11}) on the \hi\ maps at 70 and 160~$\mu$m towards the
position of the G29-UC region and G29-HMC core. The \hi\ source in our catalog is \#242. As
seen in Table~1, this is the brightest source in all 5 \hi\ bands. At
2.7 and 1.4~mm, the G29-UC region is outlined by continuum emission showing a cometary
arc shape, while the G29-HMC core emission is visible westwards in front of the arc.
The emission of the G29-HMC core is better resolved at 1.4~mm, where it shows a flattened structure. The peak of the 1.4~mm continuum emission (Beltr\'an et
al.~\cite{beltran11}), indicated with a white cross in Fig.~\ref{plateau},
coincides with the G29-HMC core. As one can see in this figure, at 70~$\mu$m, the
emission seems to be mainly associated with the G29-HMC core. In fact, the peak of
the 70~$\mu$m emission coincides with that of the 1.4~mm continuum emission. At
160~$\mu$m, the emission also seems to be more associated with the HMC than with
the \UC\ region, although in this case the peak of the \hi\ emission is located
towards the north of the G29-HMC core. The angular resolution of the \hi\ emission at
250, 350, and 500~$\mu$m is not enough to properly study with which component,
the HMC or the \UC\ region, this sub-millimeter emission is associated. 

\begin{figure}
\centerline{\includegraphics[angle=0,width=8cm]{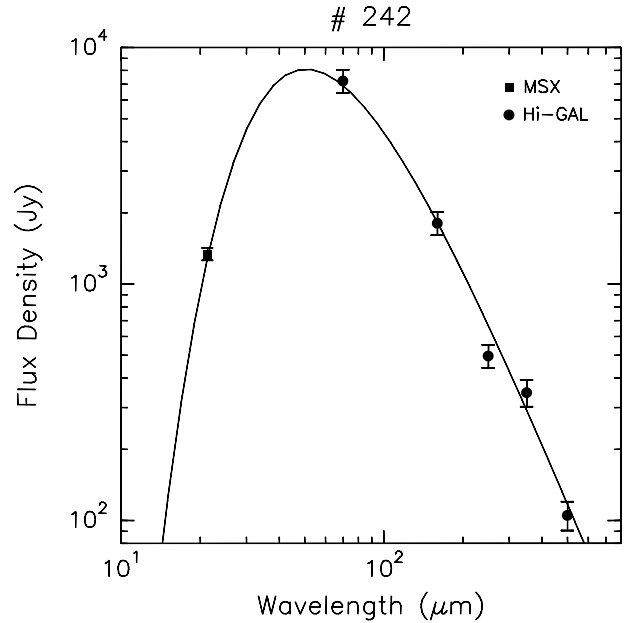}}
\caption{SED of the \hi\ source \#242, associated
with the G29-UC region and G29-HMC core. The black circles and black square
show the \hi\ and MSX at 21~$\mu$m data, respectively, with error bars. 
The black solid line represents the best-fit modified blackbody.}
\label{sed}
\end{figure}

\begin{figure}
\centerline{\includegraphics[angle=0,width=8.5cm]{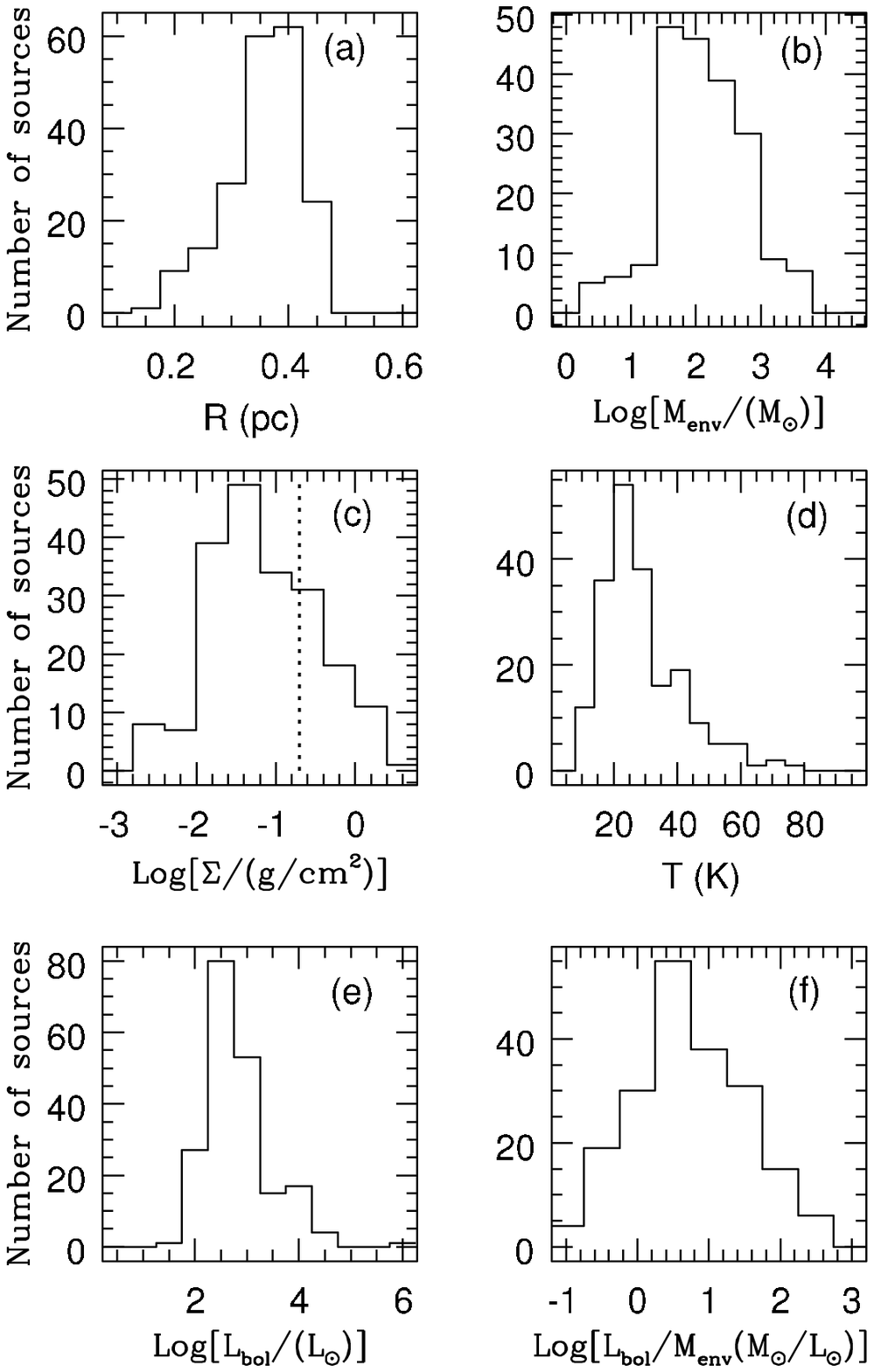}}
\caption{Histograms of some parameters of the \hi\ sources detected towards
G29: {\it a)} radius of the sources; {\it b)} mass; {\it c)} H$_2$
surface density; {\it d)} temperature; {\it e)} luminosity; and {\it f)}
luminosity-to-mass ratio. The dotted line in panel $c$ indicates the minimum surface density needed to form
massive stars according to theory (Butler \& Tan~\cite{butler12}).}
\label{histo1}
\end{figure}

From the SED fitting (Fig.~\ref{sed}), we derived a mass of $\sim$2880~$M_\odot$
for source \#242, for a dust temperature of 77~K, the highest of the sources in
the G29-SFR cloud, a size of $\sim$$19''$, and a dust emissivity index of 0.8.
The surface density is 2.3~g\,cm$^{-2}$, well above the theoretical threshold of
1~g\,cm$^{-2}$ (Krumholz \& McKee~\cite{krumholz08}) necessary for high-mass
star formation to occur. The luminosity of this source is
$\sim$$8\times10^5$~$L_\odot$ and is the highest in the whole cloud.  Kirk et
al.~(\cite{kirk10}) constructed the SED of this source by using  the SPIRE
Fourier Transform Spectrometer data from 190 to 670~$\mu$m and archival data
from 2.4 to 1.3~mm (see their Fig.~1). From the SED fitting, these authors
obtained a temperature of $\sim$80~K, in agreement with our value, and a dust
emissivity index of $\sim$1.7, twice the one that we obtained. The dust
luminosity integrated under the fitted modified blackbody in the range
2--2000~$\mu$m is $1.6\times10^6$~$L_\odot$, assuming a distance of 8.9~kpc. The
luminosity would be $\sim$$8\times10^5$~$L_\odot$ for a distance of 6.2~kpc, in
agreement with our estimated $L_{\rm bol}$. As for the mass, Kirk et
al.~(\cite{kirk10}) estimate a mass of 1500~$M_\odot$, assuming a distance of
8.9~kpc, using the fitted dust temperature and the SCUBA 850~$\mu$m flux density
(Thompson et al.~\cite{thompson06}). The mass would be $\sim$730~$M_\odot$ for a
distance of 6.2~kpc. This value is a factor $\sim$4 smaller than the one that we
obtained from the SED fitting. Besides the different method used to estimate the
mass, this difference could be accounted for, in part, by the different opacity
coefficient (0.01~cm$^2$/g at 850~$\mu$m) and dust emissivity index
($\beta$=1.7) used by these authors.

\subsection{Source physical parameters}
\label{histo-sect}

\begin{figure*}
\centerline{\includegraphics[angle=-90,width=14.5cm]{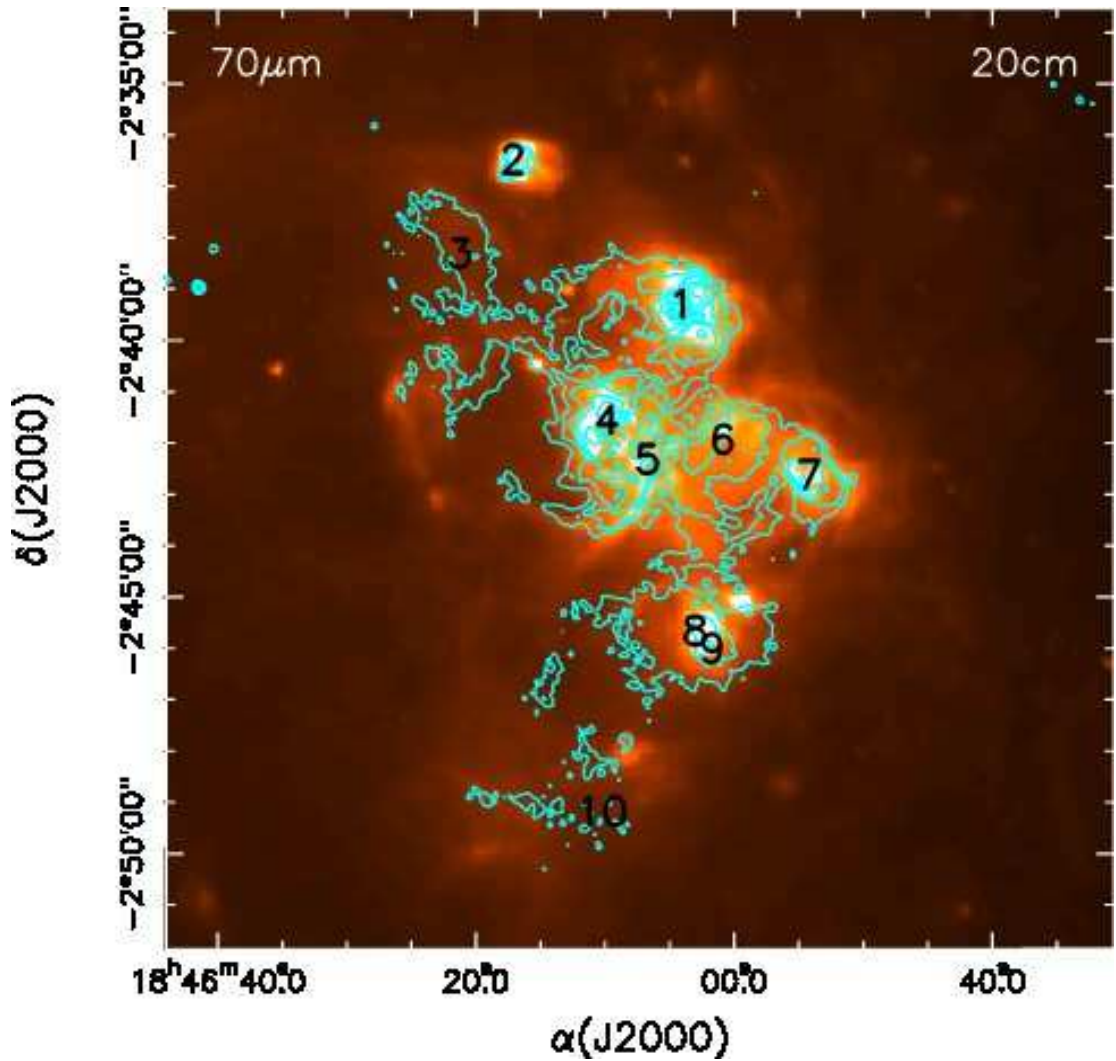}}
\caption{20~cm continuum emission from the MAGPIS survey ({\it cyan contours})
of the G29-SFR cloud overlaid on the Hi-GAL 70~$\mu$m emission ({\it colors}).
Contour levels are 3, 6, 12, 24, 48, 96, and 192 times 1~mJy\,beam$^{-1}$. The
numbers indicate the positions of the 10 NVSS sources. The synthesized beam of
the 20~cm MAGPIS observations is $6\farcs2\times 5\farcs4$ at P.A.\ = 0$\degr$}
\label{nvss}
\end{figure*}

Figure~\ref{histo1} shows the distribution of radii, masses, surface densities,
temperatures, luminosities and luminosity-to-mass ratios of the sources.
Table~\ref{taverage} shows the mean and median values for the same physical
quantities. Beltr\'an et al.~(\cite{beltran06}) observed a
sample of southern hemisphere high-mass protostellar candidates at 1.2~mm with
the SEST antenna. In the following we will
confront the physical parameters obtained for the sources in the G29-SFR cloud
with those of Beltr\'an et al.~(\cite{beltran06}) because them carried
out a detailed comparison of the values of their sources with those
estimated in other millimeter continuum surveys. The mean and median values of 0.36~pc for the radius of the \hi\ sources
associated with the G29-SFR cloud suggest that these sources are probably clumps
(e.g.\ Giannini et al.~\cite{giannini12}) that will not form individual stars
but multiple star systems or star clusters. Unfortunately, the {\it Herschel}
observations do not have enough spatial resolution to resolve these clumps into
individual cores or stars.  These values of
the radius are consistent with the mean and median values of 0.25 and 0.2~pc
found by Beltr\'an et al.~(\cite{beltran06}). The mean and median values of the mass are also consistent with the mean
and median values of 320 and 102~$M_\odot$ found by Beltr\'an et
al.~(\cite{beltran06}) for their sample, and indicates that the sources
associated with the G29-SFR cloud and detected by ${\it Herschel}$ are mostly
massive objects. The mean temperature is in agreement with the mean temperature
of 28~K found by Beltr\'an et al.~(\cite{beltran06}), and with the value of 32~K
found by Molinari et al.~(\cite{molinari00}) for a sample of luminous high-mass 
protostellar candidates in the northern hemisphere.

The average and median values of the surface density, of 0.24 and
0.06~g\,cm$^{-2}$, are similar to the mean and median values of 0.4 and
0.14~g\,cm$^{-2}$ estimated by Beltr\'an et al.~(\cite{beltran06}). These values
are slightly lower than the minimum surface density needed, according to theory
(Krumholz \& McKee~\cite{krumholz08}), to form massive stars. In a recent work,
Butler \& Tan~(\cite{butler12}) find typical mass surface densities of
0.15~g\,cm$^{-2}$ for cores, and of 0.3~g\,cm$^{-2}$ for clumps in infrared dark
clouds, some of which are likely to form massive stars. Butler \&
Tan~(\cite{butler12}) consider the cores as structures of about 100~$M_\odot$
embedded in clumps. These cores, which are virialized and in approximate
pressure equilibrium with the surrounding clump environment, are undergoing
global collapse to feed a central accretion disk. On the other hand, the clump
is defined as the gas cloud that fragments to form a star cluster. These authors
propose that fragmentation in these clumps could be inhibited by magnetic fields
rather than radiative heating and that the initial conditions of local massive
star formation in the Galaxy may be better characterized by surface density
values of $\sim$0.2~g\,cm$^{-2}$ rather than 1~g\,cm$^{-2}$. This would imply
smaller accretion rates and longer formation timescales ($> 10^5$~yr) for massive stars than
those predicted my McKee \& Tan~(\cite{mckee03}).

The mean luminosity estimated, $6.2\times10^3$~$L_\odot$, would correspond to a
main-sequence star of spectral type B1 following Table~1 of Mottram et
al.~(\cite{mottram11}), and thus, it also indicates that the \hi\ sources
associated with the G29-SFR cloud are mostly high-mass sources. Note, however,
that this value is an order of magnitude smaller than the average value of
$6.7\times10^4$~$L_\odot$ obtained by Beltr\'an et al.~(\cite{beltran06}) for a
sample of massive protostellar candidates. This is not surprising, taking into
account that the bolometric luminosities calculated by these authors are to be
considered upper limits because estimated from the IRAS flux densities. The IRAS
beam is so large ($\sim$2$'$) that when integrating the flux density for a
single protostellar candidate,  there might be an important contribution not
only from other sources that may fall into such a large beam, but also from
inter-clump diffuse emission. The latter contribution is subtracted out when
doing the source extraction but is included if one simulates what would be seen
with a larger beam like that of IRAS.

The luminosity-to-mass ratio, $L_{\rm bol}/M_{\rm env}$, is an important parameter for establishing
the age of a source. This ratio is expected to increase with time as more gas is
incorporated into the star that becomes more luminous. The mean and median
$L_{\rm bol}/M_{\rm env}$
values for the sources in the G29-SFR cloud are 23 and 5~$L_\odot/M_\odot$,
respectively, which are significantly lower than the average and median values
of 99~$L_\odot/M_\odot$ obtained by Beltr\'an et al.~(\cite{beltran06}).
However, as already mentioned, this discrepancy could be due to the fact that
the bolometric luminosities of the sources in the Beltr\'an et al.\ sample are
likely upper limits because they were estimated from the IRAS fluxes.

\begin{figure*}
\centerline{\includegraphics[angle=-90,width=15cm]{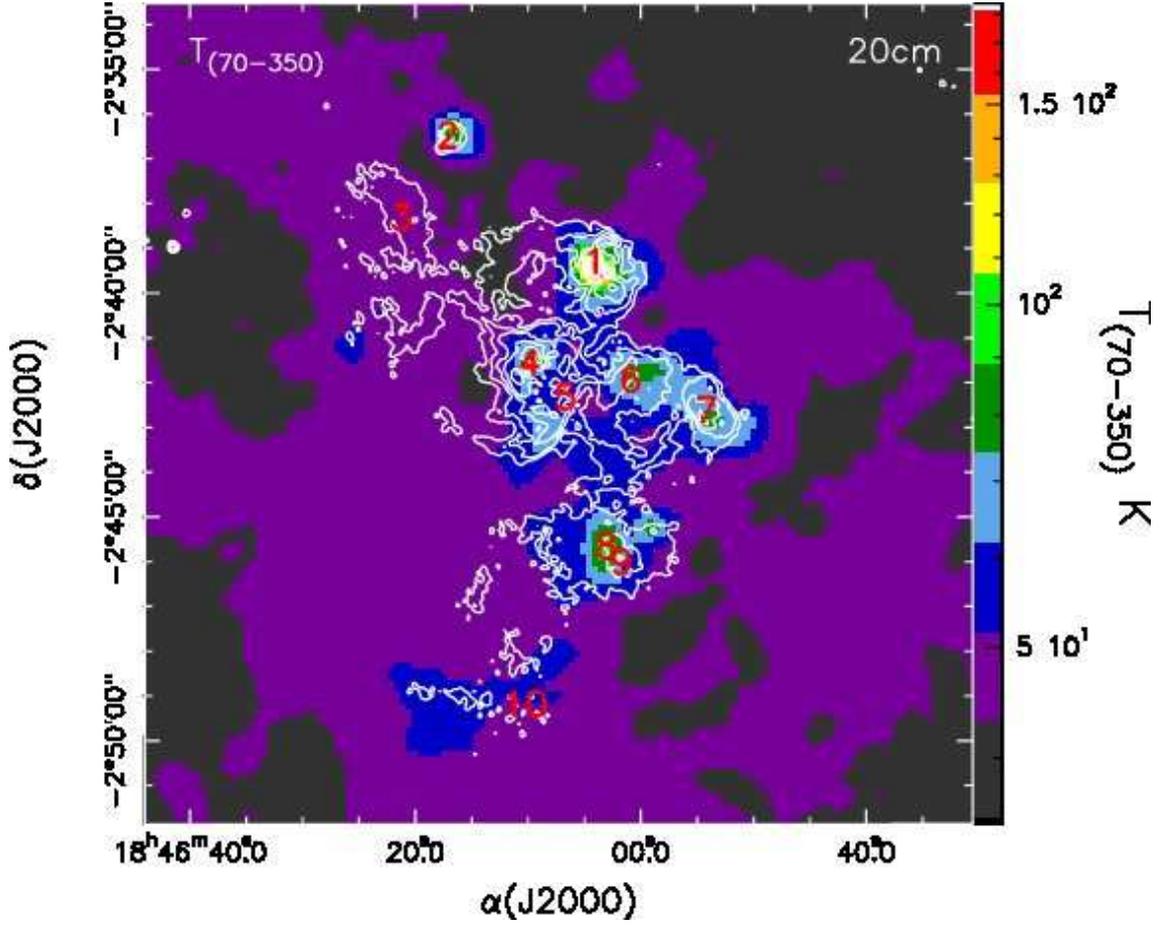}}
\caption{20~cm continuum emission from the MAGPIS survey ({\it blue contours})
of the G29-SFR cloud overlaid on the 70--350~$\mu$m color temperature map in
logarithmic scale
({\it colors}).
Contour levels and numbers are the same of Fig.~\ref{nvss}.}
\label{temp-cm}
\end{figure*}

\subsection{Centimeter emission associated with the G29-SFR cloud}
\label{cm}

Figure~\ref{nvss} shows a zoom-in towards the central region of the G29-SFR
cloud. In this figure, the 20~cm emission of the  Multi-Array Galactic Plane
Imaging Survey (MAGPIS: Helfand et al.~\cite{helfand06}) is overlaid on the
Hi-GAL 70~$\mu$m  emission. The angular resolution of both sets of data is
similar, which makes the comparison straightforward. The positions of the ten
21-cm sources  associated with the cloud  from the NRAO/VLA Sky Survey (NVSS:
Condon et al.~\cite{condon98}) catalog at 1.4~GHz are also indicated in the
figure. NVSS source \#1 is associated with the G29-UC region and with \hi\
source \#242. Table~\ref{tnvss} gives the coordinates, flux densities at 21~cm,
and major and minor axes of the NVSS sources after deconvolving with the
restoring beam of 45$''$ of the NVSS images. The source fluxes have been
obtained from the NVSS catalog instead of estimating them directly from the
MAGPIS map at 20~cm because of the better surface brightness sensitivity of
NVSS. The deconvolved sizes for five out of ten sources are upper limits, which
indicates that either the source is unresolved or that the emission is too large
to properly fit it with just one Gaussian (Condon et al.~\cite{condon98}). The
latter is the case for NVSS source \#10, which, as seen in Fig.~\ref{nvss}, is
very extended and with a very low-level emission, and therefore difficult to fit
with a Gaussian. Figure~\ref{temp-cm} shows the 20~cm continuum emission
overlaid on the 70--350~$\mu$m color temperature, $T_{(70-350)}$, map. To
calculate the color-color  temperature map, we first smoothed the 70~$\mu$m map
(that with the highest angular resolution: $9\farcs2$) to the resolution of the
350~$\mu$m map ($25''$), and then reprojected both maps to the same pixel and
map size. These two wavelengths happen to bracket the peak of the SED and are
hence most sensitive to temperature changes. Clearly, the colour temperature is
a proxy for the dust temperature, but may differ significantly from the
temperature estimate obtained by fitting the whole SED with a modified
blackbody. As seen in Fig.~\ref{temp-cm}, the positions of the NVSS sources,
except for the very diffuse NVSS sources \#3 and 10, coincide with local maxima
of the color-color temperature.

\begin{table*}
\caption[] {Position, fluxes, and sizes of the
NVSS sources associated with the G29-SFR cloud} 
\label{tnvss}
\begin{tabular}{rccccc}
\hline
&\multicolumn{1}{c}{$\alpha({\rm J2000})$} &
\multicolumn{1}{c}{$\delta({\rm J2000})$} &
\multicolumn{1}{c}{$S_{21\,{\rm cm}}$} &
\multicolumn{1}{c}{Major axis} & 
\multicolumn{1}{c}{Minor axis} 
\\
\multicolumn{1}{c}{\# Id.} &
\multicolumn{1}{c}{($^{\rm h}$ $^{\rm m}$ $^{\rm s}$)}&
\multicolumn{1}{c}{($\degr$ $\arcmin$ $\arcsec$)} &
\multicolumn{1}{c}{(Jy)} &
\multicolumn{1}{c}{($''$)} &
\multicolumn{1}{c}{($''$)} 
\\  
\hline
 1$^a$ & 18 46 04.09 & $-$2 39 19.1  & 2.38    &    26.6  &    20.4  \\
 2 & 18 46 17.11 & $-$2 36 30.0  & 0.025   & $<$26.3  & $<$18.4  \\
 3 & 18 46 21.24 & $-$2 38 20.3  & 0.035   &    63.5  &    31.2  \\ 
 4 & 18 46 09.83 & $-$2 41 34.9  & 3.06    &    49.3  &    44.7  \\
 5 & 18 46 06.69 & $-$2 42 20.8  & 1.55    &   127.3  &    50.1  \\
 6 & 18 46 00.89 & $-$2 41 57.4  & 0.137   & $<$14.2  & $<$14.1  \\
 7 & 18 45 54.17 & $-$2 42 39.0  & 0.435   &    35.6  &    24.5  \\
 8 & 18 46 03.00 & $-$2 45 41.0  & 0.097   & $<$80.3  & $<$21.6  \\
 9 & 18 46 01.67 & $-$2 46 01.6  & 0.052   & $<$15.5  & $<$15.3  \\
10 & 18 46 10.25 & $-$2 49 12.3  & 0.010   &$<$125.9  &$<$125.0  \\
\hline
\end{tabular}
\\
$^a$ G29-UC and \hi\ source \#242.
\end{table*}

\begin{table*}
\caption[] {Physical parameters of the \HII regions in the G29-SFR cloud} 
\label{tparam}
\begin{tabular}{rcccccccc}
\hline
&\multicolumn{1}{c}{$R$} &
\multicolumn{1}{c}{$T_B$} &
\multicolumn{1}{c}{$n_e$} &
\multicolumn{1}{c}{$EM$} &
\multicolumn{1}{c}{$N_{\rm Ly}$} & 
\multicolumn{1}{c}{$M_{\rm ion}$} &
\multicolumn{1}{c}{Spectral} 
\\
\multicolumn{1}{c}{\# Id.} &
\multicolumn{1}{c}{(pc)}&
\multicolumn{1}{c}{(K)}&
\multicolumn{1}{c}{(cm$^{-3}$)} &
\multicolumn{1}{c}{($10^5$ cm$^{-6}$ pc)} &
\multicolumn{1}{c}{($10^{47}$ s$^{-1}$)} &
\multicolumn{1}{c}{($M_\odot$)}&
\multicolumn{1}{c}{Type} 
\\  
\hline
 1$^a$ & 0.35    &274    & 2028    &  19     & 84	&9.0     & O6 \\
 2 &$<$0.33  &$>$32  &$>$226   &$>$0.23  & 0.88	&$<$0.85 & B0   \\
 3 & 0.67    &11     &  93     &  0.08   & 1.2	&2.9     & B0   \\ 
 4 & 0.71    &865    & 803     &  6      & 108	&29      & O5   \\
 5 & 1.2     &152    & 258     &  1      & 55	&46      & O6.5 \\
 6 &$<$0.21  &$>$426 &$>$1027  &$>$3     & 4.8	&$<$1.0  & O9.5 \\
 7 & 0.44    &311    & 607     &  2      & 15	&5.5     & O8.5 \\
 8 &$<$0.63  &$>$35  &$>$171   &$>$0.24  & 3.4	&$<$4.3  & 09.5 \\
 9 &$<$0.23  &$>$136 &$>$557   &$>$0.96  & 1.8	&$<$0.71 & B0   \\
10 &$<$1.9   &$>$0.4 &$>$ 10   &$>$0.003 & 0.35	&$<$7.2  & B0.5 \\
\hline
\end{tabular}
\\
$^a$ G29-UC and \hi\ source \#242.
\end{table*}

As seen in Fig.~\ref{nvss}, the centimeter emission is well correlated with the
70~$\mu$m emission, even at the low level emission. Note how both the centimeter
and the FIR emission trace the arcs seen eastwards of NVSS sources \#4 and 5. 
These arcs are shock fronts where hydrogen is ionized, and gives rise to the
radio continuum. It is also possible that important shock gas coolants like the
[OI 63~$\mu$m] line could be in part contaminating the PACS\,70~$\mu$m emission. The fact that the centimeter emission is so extended and well correlated with
the dust emission would suggest that it is associated with a group of \HII
regions that are ionizing and disrupting the cloud. Assuming that the centimeter
continuum emission comes from homogeneous optically thin \HII regions, we
calculated the physical parameters of the 10 NVSS sources (using the formalism
of Mezger \& Henderson~\cite{mezger67} and Rubin~\cite{rubin68}) and list them
in Table~\ref{tparam}. Column 1 gives the NVSS number of the source
(Table~\ref{tnvss}), column 2 the spatial radius $R$ of the \HII region, which
was determined from the deconvolved source size (Table~\ref{tnvss}), column 3
the source averaged brightness $T_B$, column 4 the  electron density $n_e$,
column 5 the emission measure $EM$, column 6 the number of Lyman-continuum
photons per second $N_{\rm Ly}$, column 7 the  mass of ionized gas $M_{\rm
ion}$, which was calculated assuming a spherical homogeneous distribution, and
column 8 the spectral type of the ionizing source. The spectral type was
computed from the estimated $N_{\rm Ly}$ and using the tables of Davies et
al.~(\cite{davies11}) and Mottram et al.~(\cite{mottram11}), which are for 
Zero Age Main Sequence (ZAMS) stars. Note that if the 21~cm emission is optically thick, then $T_B$, $n_e$, $EM$, $N_{\rm
Ly}$,  $M_{\rm ion}$, and therefore, the spectral type should be considered as
lower limits. For the sources with upper limits for the deconvolved sizes
(Table~\ref{tnvss}), $R$ and $M_{\rm ion}$ should be taken as upper limits,
while $T_B$, $n_e$ and $EM$ as lower limits. As seen in Table~\ref{tparam}, most
of the sources are early B or late O types. However, the cloud would also
contain 3 sources, with one of them being the G29-UC region (NVSS source \#1), with spectral types
O5--O6.5. Therefore, it is possible that these massive sources, with their
strong winds and radiation pressure, are disrupting and shaping the cloud. This
effect may contribute to underestimate the number of ionizing photons and, in turn,
the luminosities of the stars. Note that sources \#4 and 5, located at the head
of the large arc-like structure seen towards the east, have spectral types O5
and O6.5, respectively.

\subsection{Physical parameters as a function of the distance to the NVSS
sources in the G29-SFR cloud}

\begin{figure}
\centerline{\includegraphics[angle=0,width=8.5cm]{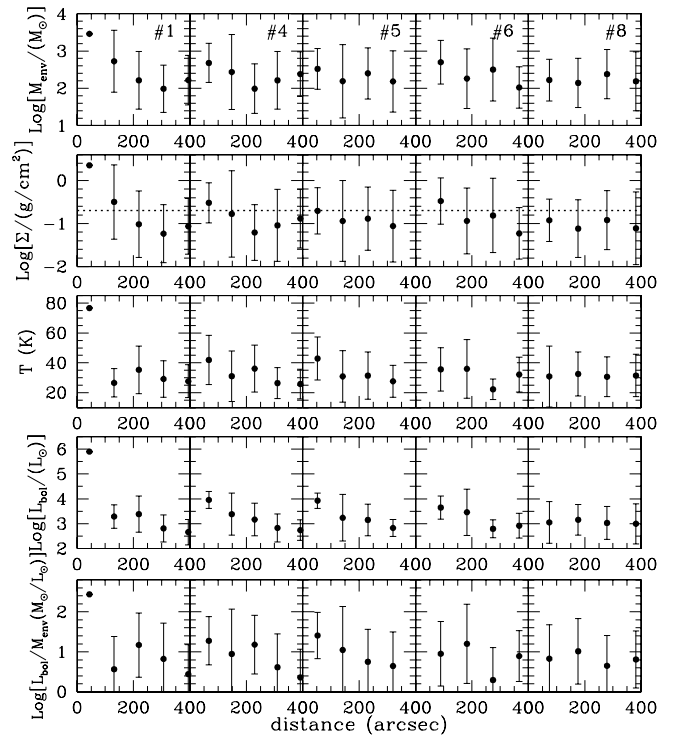}}
\caption{Distributions of mass, surface density, temperature, luminosity, and 
luminosity-to-mass ratio as a function of the distance to the NVSS sources in
the G29-SFR cloud. The NVSS number (Tables~\ref{tnvss}
and \ref{tparam}) is indicated in the lefthand upper
corner of the upper panels. The dotted line in the surface density 
distributions indicates the minimum value needed needed to form
massive stars according to theory (Butler \& Tan~\cite{butler12}). The first bin
contains only one point and, thus, the standard deviation is zero.}
\label{dg29}
\end{figure}

To check whether there is a variation of the \hi\ source physical parameters as a
function of the distance to the most massive sources in the G29-SFR cloud, we
plotted the distribution of masses, surface densities, luminosities, 
temperatures, and luminosity-to-mass ratios of the \hi\ sources as a function of
the distance to the NVSS sources \#1, 4, 5, 6, and 8 (Fig.~\ref{dg29}). The NVSS sources \#2, 3, 7, and 10 have not been taken into account because are
located close to the border of the cloud. NVSS source \#9 is located very close
to NVSS source \#8, and therefore, the distributions should be very similar. The
data have been binned in intervals of $\sim$80$''$.  For NVSS source \#1
(G29-UC), only one \hi\ source  (\#242) is found in the first interval of
$\sim$80$''$,  which means that the first point in the plots takes into account
only the physical parameters of this source. 

The physical parameters of NVSS source \#1 (G29-UC) and its immediate
surroundings have the highest values of all the centimeter sources in the
G29-SFR cloud. This is evident in Fig.~\ref{dg29} when comparing NVSS source \#1
to sources \#4, 5, 6, and 8, but it is also true for the rest of NVSS sources
not shown in this plot. Given the large error bars in Fig.~\ref{dg29}, one
can only see a marginal trend of the mass and surface density, which seem to decrease with the
distance from NVSS source \#1. The surface density is above the minimum
value of 0.2~g\,cm$^{-2}$ needed to form massive stars according to theory
(Butler \& Tan~\cite{butler12}), up to a distance of $\sim$150$''$ from the
G29-UC region. A similar marginal trend is seen for NVSS source \#4.
Although in this case, the decrease in $M_{\rm env}$ is even less obvious,
and the surface density is slightly above 0.2~g\,cm$^{-2}$ only in a small
region ($\lesssim80''$) surrounding the source. Regarding the luminosity,
temperature and luminosity-to-mass ratio, again, the highest values are
found towards the NVSS source \#1 (G29-UC).

The fact that the most massive and luminous \hi\ sources in the cloud are
located close to the strongest source in the G29-SFR cloud (\#242 or G29-UC)
suggest that there is a privileged area for massive star formation in the cloud.
Based on the central location of the G29-UC region inside the G29-SFR cloud
(Fig.~\ref{co-sou}), this indicates that high-mass stars form preferentially
at the center of the cloud, as expected. An inhomogeneous density distribution of the cloud,
with higher density towards the center of the cloud (maybe already present as an
initial condition), could be responsible for this source distribution. This
is consistent with the findings of most millimeter continuum surveys.

\subsection{24$\mu$m-dark versus 24$\mu$m-bright sources}
\label{preproto}

Because star formation does not occur simultaneously all over a cloud, one would
expect to find young stellar objects in different evolutionary stages associated
with the G29-SFR cloud. To search for differences in the evolutionary stage of the sources, we
cross-correlated our \hi\ sources with the {\it Spitzer} MIPSGAL 24~$\mu$m
catalog. The last column in Table~1 indicates whether a source is associated or
not with 24~$\mu$m emission. Obviously, we counted as associated those sources
saturated at 24~$\mu$m, like for example the \hi\ source \#242 (G29-UC). Based
on this association, we divided the sources into two groups: those without a
24~$\mu$m counterpart, that we call 24$\mu$m-dark, and those with a 24~$\mu$m
counterpart, that we call 24$\mu$m-bright. The former are expected to be the
youngest \hi\ sources in the cloud. As a result of this cross-correlation we
discovered 81 \hi\ sources not associated with 24~$\mu$m emission and  117 \hi\
sources associated with it. As shown in Fig.~\ref{co-sou}, both kind of Young
Stellar Objects (YSOs) are uniformly distributed over the cloud.

\begin{figure}
\centerline{\includegraphics[angle=0,width=8.5cm]{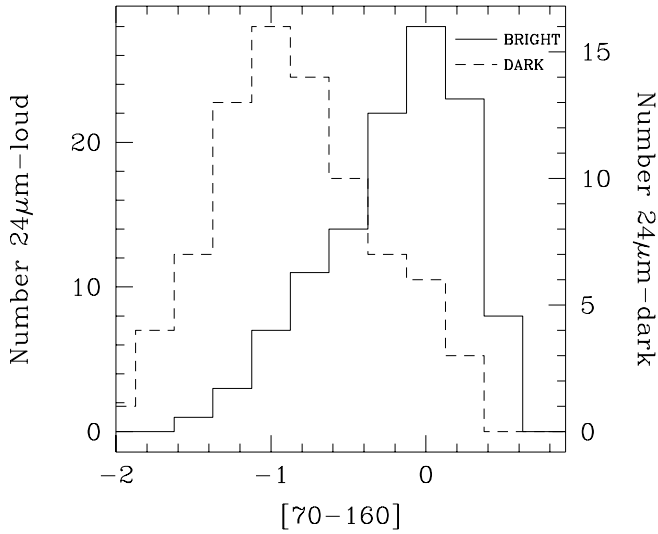}}
\caption{Histogram of the [70--160] color for 24~$\mu$m-bright ({\it solid line})
and 24~$\mu$m-dark ({\it dashed line}) sources.}
\label{70-160}
\end{figure}

All the sources in our sample have been selected to be detected in all 5
photometric \hi\ bands. Therefore, by definition, all the sources have been
detected at 70~$\mu$m, which would suggest that most of them, if not all, are
protostellar. However, this does not mean that there are no prestellar sources
in the G29-SFR cloud (see Pillai et al.~\cite{pillai11}). The analysis of the
sources not detected at 70~$\mu$m, and likely prestellar, although being highly
interesting, goes beyond the scope of the present study. To check whether 
24$\mu$m-dark and 24$\mu$m-bright sources show any difference in their 70~$\mu$m
fluxes, we plotted and histogram of the [70--160] color for both kind of 
sources (Fig.~\ref{70-160}). As seen in this figure, the [70--160] color of  
24$\mu$m-dark sources is clearly smaller than those of the 24$\mu$m-bright ones.
This indicates that the possible different evolutionary phase of the sources is
also supported by the \hi\ data.

\begin{figure}
\centerline{\includegraphics[angle=0,width=9cm]{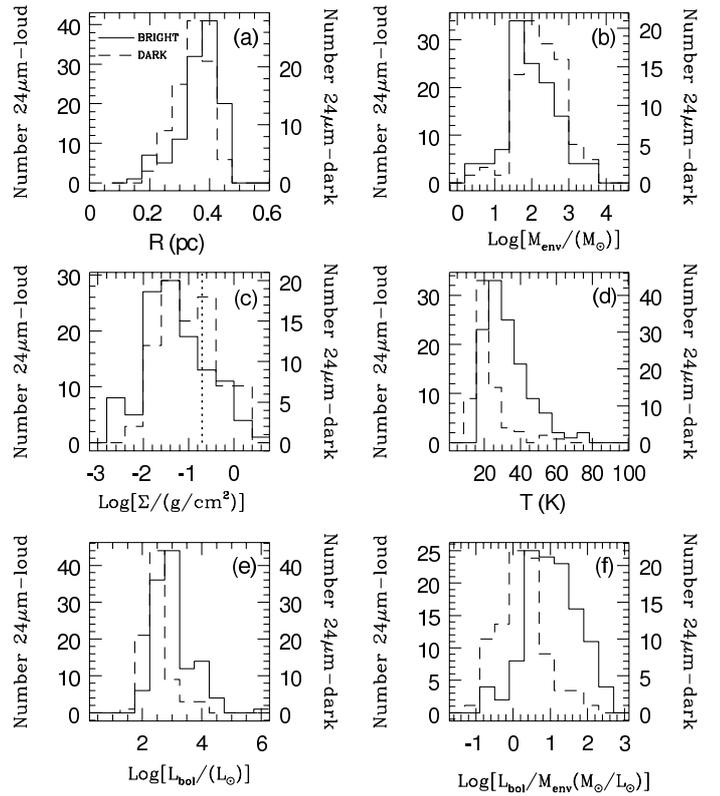}}
\caption{Histograms of {\it a)} radius of the sources; {\it b)} mass; {\it c)} H$_2$
surface density; {\it d)} temperature; {\it e)} luminosity; and {\it f)}
luminosity-to-mass ratio, for 24~$\mu$m-bright 
({\it solid line}) and 24~$\mu$m-dark ({\it dashed line}) sources. The dotted line in panel $c$ indicates the minimum surface density needed to form
massive stars according to theory (Butler \& Tan~\cite{butler12}).}
\label{mips}
\end{figure}

Figure~\ref{mips} shows the distribution of radii, masses, surface densities,
temperatures, luminosities and luminosity-to-mass ratios for 24$\mu$m-dark and
24$\mu$m-bright sources. Table~\ref{taverage} shows the mean and median values for
the same physical quantities. One sees that the distributions of the two types
of objects are different.  A closer inspection of the data using the
Kolmogorov-Smirnov (KS) statistical test shows that, except for the radius distributions, 
the probability of the mass, surface density, temperature, luminosity, and
luminosity-to-mass ratio distributions being the same for 24$\mu$m-dark and
24$\mu$m-bright sources is very low ($P$ $\lesssim$ 0.004). Therefore, the
physical properties of the two groups are statistically different. The temperature, luminosity, and, in particular, the luminosity-to-mass ratio are
smaller for the 24$\mu$m-dark than for the 24$\mu$m-bright  objects, while the
mass and the surface density are higher.  That $T_d$, $L_{\rm bol}$, and $L_{\rm
bol}/M_{\rm env}$ are smaller for 24$\mu$m-dark  than for 24$\mu$m-bright sources
is consistent with the former being in an earlier evolutionary phase. 
Figure~\ref{mips} also shows that a relatively large number of 24$\mu$m-dark and
24$\mu$m-bright sources have surface densities high enough to form massive stars
according to theory (Butler \& Tan~\cite{butler12}).

The most significant difference between the two groups is found in the value of
$L_{\rm bol}/M_{\rm env}$. In fact, the mean and median value of $L_{\rm bol}/M_{\rm env}$ is $\sim$6 and $\sim$5 times
lower for the 24~$\mu$m-dark sources compared to the 24~$\mu$m-bright ones,
which supports our assumption that the sources not associated with 24~$\mu$m
emission are in an earlier evolutionary phase.

\begin{figure}
\vspace*{-1.1cm}
\centerline{\includegraphics[angle=0,width=13.5cm]{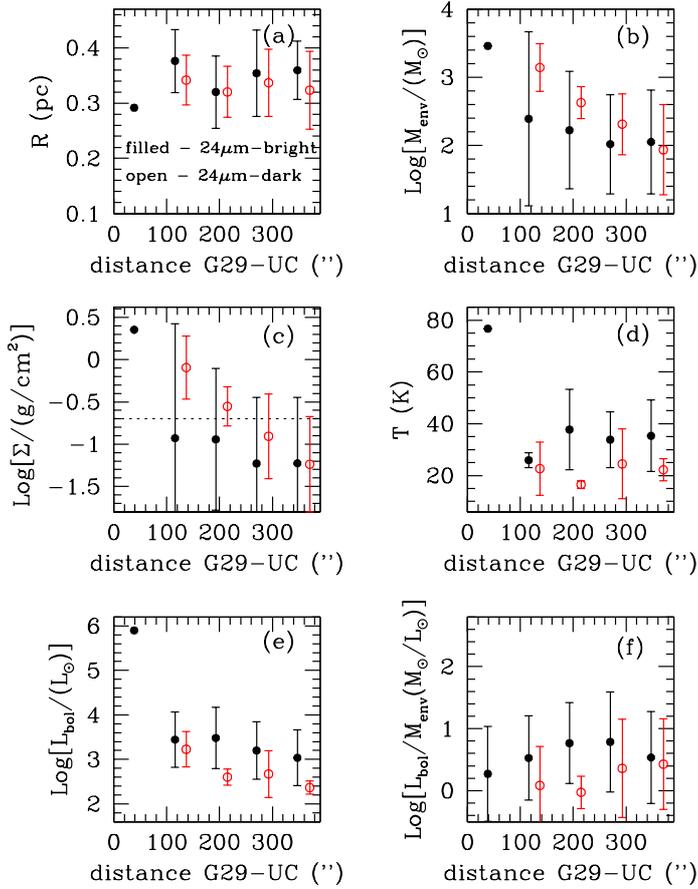}}
\caption{Distributions of {\it a)} radius of the sources; {\it b)} mass; {\it c)} H$_2$
surface density; {\it d)} temperature; {\it e)} luminosity; and {\it f)}
luminosity-to-mass ratio as a function of the distance to the G29-UC region
for 24~$\mu$m-bright ({\it filled circles}) and 24~$\mu$m-dark ({\it open red circles})
sources. The dotted line in 
the surface density distributions indicates the minimum value needed to form
massive stars according to theory (Butler \& Tan~\cite{butler12}). The first bin
contains only one point and, thus, the standard deviation is zero.}
\label{dg29-pre}
\end{figure}

We also investigated whether the radii, masses, surface densities, temperatures,
luminosities and luminosity-to-mass ratios of the two types of sources show any
correlation as a function of the distance to the G29-UC region.
Figure~\ref{dg29-pre} indicates that both groups show the same trends, that is,
the mass, surface density, and luminosity of the sources marginally decrease
when moving away from the G29-UC region, while the size, temperature and
luminosity-to-mass ratio, except for the high values close to the G29-UC region,
do no significantly change. 

\section{Discussion}

\subsection{Evolutionary phase of the sources}
\label{evol}

To investigate the stability of the sources, we calculated their Jeans masses,
$M_J$, and virial masses, $M_{\rm virial}$. $M_J$ was calculated following the
expression  $M_J = [T/10\,{\rm K}]^{3/2}\times[n_{\rm H_2}/1\times10^4\,{\rm
cm^{-3}}]^{-1/2}$, where the dust temperature $T$ was obtained from the SED fitting
and the H$_2$ volume density $n_{\rm H_2}$ was calculated assuming that the sources
have spherical symmetry (the size of the sources is that obtained from the source extraction
process). $M_{\rm virial}$ was estimated from the line width,
$\Delta V$, of $^{13}$CO~(1--0) towards the position of each source following
the expression of  MacLaren et al.~(\cite{macLaren88}), $M_{\rm virial} =
0.509\times d\times \theta\times \Delta V^2$,  where $d$ is the distance in kpc,
$\theta$ is the size of the source in arcsec obtained from the source extraction
process, and $\Delta V$ is in \kms. The choice of $^{13}$CO~(1--0) to estimate
the virial masses, which could be partially optically thick and therefore
overestimate the line width, is based on the fact that it is the only molecular
tracer covering the whole cloud. To have an idea of how large the overestimate
of the line widths could be, we checked the value towards source \#242 (associated
with the G29-UC region and G29-HMC core), which has been extensively observed in different molecular tracers. The line
width estimated with $^{13}$CO is 6.2~\kms\ and is similar to the values of
$\sim$5.5~\kms\ estimated in CS~(5--4) and (7--6), and HCO$^+$~(3--2) with the
JCMT and the IRAM 30-m telescopes (Olmi et al.~\cite{olmi99}; Churchwell et
al.~\cite{churchwell10}). $M_{\rm virial}$ depends on the density profile, and
for a power-law density distribution of the type $\rho \propto r^{-p}$, the
virial mass should be multiplied by a factor $3(5-2p)/5(3-p)$, which is $\leq 1$
for $p< 3$. Thus, the values estimated should be taken as upper limits. 

\begin{figure}
\centerline{\includegraphics[angle=0,width=7.5cm]{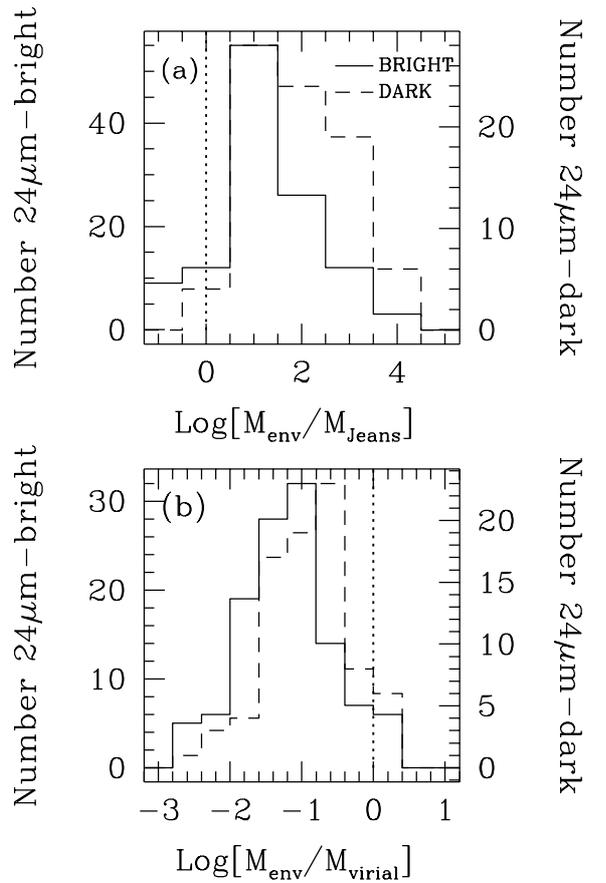}}
\caption{Histogram of {\it a)} the mass-to-Jeans mass ratio and {\it b)}
mass-to-virial mass ratio for 24~$\mu$m-bright 
({\it solid line}) and  24~$\mu$m-dark ({\it dashed line}) sources. The
dotted vertical line indicates  {\it a)} $M_{\rm env}=M_J$ and  {\it b)} $M_{\rm
env}=M_{\rm virial}$.}
\label{jeans}
\end{figure}

Figure~\ref{jeans} shows the $M_{\rm env}$--$M_J$ ratio and the $M_{\rm
env}$--$M_{\rm virial}$ ratio for all the sources, 24~$\mu$m-bright and
24~$\mu$m-dark. As seen in this plot, almost all the sources have masses well
above $M_J$.  In particular, 90\% of the 24~$\mu$m-bright sources and 96\% of
the 24~$\mu$m-dark ones have masses well above $M_J$. In fact, the mean and
median values of the $M_{\rm env}$--$M_J$ ratio are 296 and 14 for
24~$\mu$m-bright sources, and 735 and 86 for 24~$\mu$m-dark sources.  This
indicates that most of the sources in the G29-SFR cloud would be gravitationally
supercritical if only supported by thermal pressure, in which case, they should
be collapsing. The $M_{\rm env}$--$M_{\rm virial}$ ratio confirms that an
additional supporting agent, such as turbulence, is likely acting against
gravity in these sources, because only 5\% (6 out of 117 sources) of the
24~$\mu$m-bright sources and 7\% (6 out of 81 sources) of the 24~$\mu$m-dark
ones have masses above the virial mass. The mean and median values of the
$M_{\rm env}$--$M_{\rm virial}$ ratio are 0.2 and 0.07 for 24~$\mu$m-bright
sources, and 0.3 and 0.1 for 24~$\mu$m-dark sources.

This result seems to be in contrast with the results of other studies of
high-mass star-forming clumps, where the mass of the clumps is found to be
larger than the virial mass (e.g.\ Hofner et al.~\cite{hofner00}; Fontani et
al.~\cite{fontani02}). L\'opez-Sepulcre et al.~(\cite{lopez10}) findings  for a
sample of 29 IR-bright and 19 IR-dark high-mass cluster-forming clumps are
similar to ours, although on average their objects are closer to virial
equilibrium.  What are the sources of uncertainty in our estimate of the $M_{\rm
env}$--$M_{\rm virial}$ ratio?  The major problem is that very likely the
$^{13}$CO emission is not tracing the same volume of gas as the 1.2~mm continuum
emission. This means that the $^{13}$CO line width may not be representative of
the gas contributing to $M_{\rm env}$. However, to allow for a mean value of
$M_{\rm env}/M_{\rm virial}$$=1$, one should shift the distributions in Fig. 12b
by an order of magnitude, which implies a decrease of the line width by a factor
$\sim$3. This seems too much, as observations of different tracers with
different resolutions in high-mass star forming regions reveal changes by only a
few km/s, for line widths of several km/s. Another source of error could be the
temperature estimate, which enters almost linearly into the calculation of
$M_{\rm env}$. It is thus difficult to believe that this effect may contribute
by more than 20--30\%, by far less than the factor 10 required to match  $M_{\rm
env}$ to $M_{\rm virial}$. Finally, density gradients may affect the estimate of
$M_{\rm virial}$. Assuming a power-law density profile as steep as $\rho\propto R^{-2}$,
with $R$ radius of the clump, our values of $M_{\rm virial}$ should decrease only by a
factor 0.6 (see MacLaren et al. 1988), still not sufficient to justify the
observed ratio $M_{\rm env}$--$M_{\rm virial}$.

We conclude that none of the previous effects can explain the distributions in
Fig.~12b. However, it is possible that {\it all} of them contribute to the
result. While this is certainly possible for a limited number of sources
(especially those with $M_{\rm env}/M_{\rm virial}\lesssim 1$), it seems likely
that $M_{\rm env}/M_{\rm virial}$ is indeed $<$1 for the majority of the
objects.

Assuming that this is the case, it is interesting to note that of the 36 sources located at $\lesssim 4'$ of the G29-UC region, 14\% (5 sources
including source \#242: G29-UC) have $M_{\rm env} > M_{\rm virial}$. On the
other hand, of the remaining 162 sources, which are located at $> 4'$, only 4\%
(7 sources), have $M_{\rm env} > M_{\rm virial}$. Despite the poor statistics,
this result seems to suggest that the sources that should be undergoing collapse
and forming stars are preferentially concentrated towards the dominant
source in the G29-UC cloud.

\begin{figure}
\centerline{\includegraphics[angle=0,width=9cm]{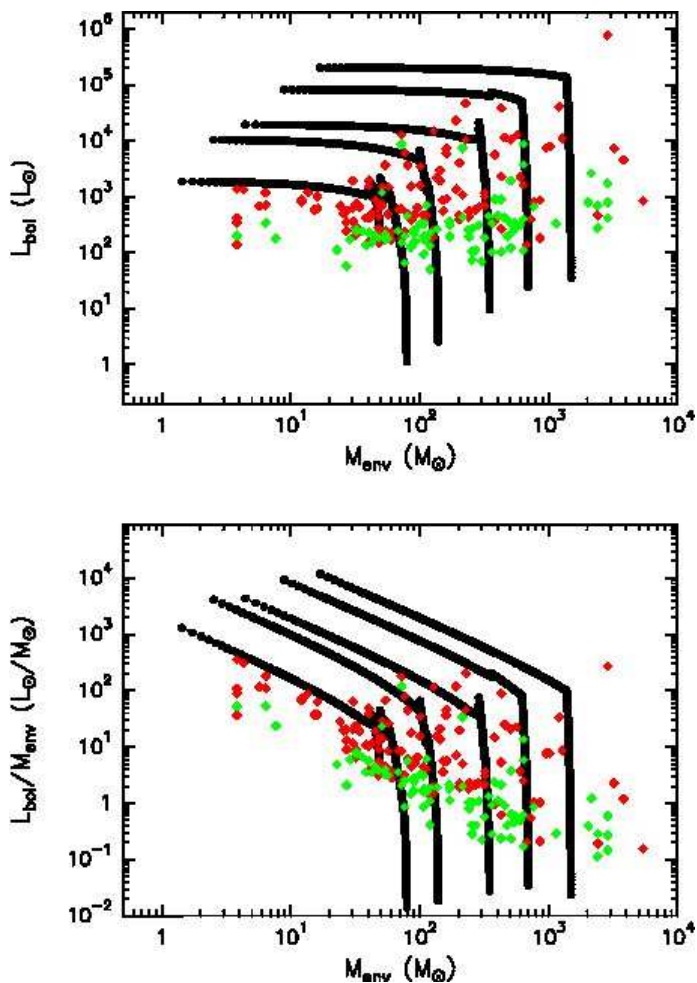}}
\caption{$L_{\rm bol}$--$M_{\rm env}$ ({\it upper panel}) and 
$L_{\rm bol}$/$M_{\rm env}$--$M_{\rm env}$ ({\it lower panel}) plots for 24~$\mu$m-bright
({\it red diamonds}) and 24~$\mu$m-dark
({\it green diamonds}) sources in the G29-SFR cloud.
Black lines represent the evolutionary tracks of Molinari et
al.~(\cite{molinari08}) (see \S~\ref{evol}). The different models 
are for different initial masses of 80, 140, 350, 700 and 
1500~$M_\odot$ (from left to right).}
\label{tracks}
\end{figure}

The fact that the sources associated with the G29-SFR cloud appear to be in
different evolutionary stages is also suggested by the association or not with
{\it Spitzer} 24~$\mu$m emission, as already discussed in \S~\ref{preproto}. 

To check the validity of this evolutionary phase difference for the sources in
the G29-SFR cloud, we decided to use the evolutionary sequence tool of Molinari
et al.~(\cite{molinari08}). These authors have developed an empirical model to
describe  the pre-main sequence evolution of YSOs in the high-mass regime based
on an $L_{\rm bol}$--$M_{\rm env}$ diagram, where $L_{\rm bol}$ is the bolometric
luminosity of the sources, and $M_{\rm env}$ the total envelope mass. Based on
the model of collapse in turbulence supported cores of McKee \&
Tan~(\cite{mckee03}), which describes the free-fall accretion of material onto a
central source as a time-dependent process, Molinari et al.~(\cite{molinari08})
have constructed evolutionary tracks in the $L_{\rm bol}$--$M_{\rm env}$ diagram.
According to this evolutionary sequence, sources in different phases should
occupy different regions of the $L_{\rm bol}$--$M_{\rm env}$ diagram. For the
high-mass regime, the bolometric luminosity of a YSO evolving towards the ZAMS
increases by several orders of magnitude during the accretion phase. Therefore,
one would expect 24~$\mu$m-dark sources to have a lower $L_{\rm bol}$ than the
24~$\mu$m-bright ones for similar $M_{\rm env}$.  Elia et al.~(\cite{elia10})
prefer to use the $L_{\rm bol}/M_{\rm env}$ ratio versus $M_{\rm env}$ as a
diagnostic, based on the fact that an earlier evolutionary stage source should
have smaller $L_{\rm bol}/M_{\rm env}$ ratio than more evolved ones.

As seen in Fig.~\ref{tracks}, 24~$\mu$m-dark and 24~$\mu$m-bright sources occupy
different regions of the $L_{\rm bol}$--$M_{\rm env}$ and $L_{\rm bol}$/$M_{\rm
env}$--$M_{\rm env}$ diagrams, with 24~$\mu$m-dark sources having lower $L_{\rm
bol}$ and $L_{\rm bol}$/$M_{\rm env}$ for similar $M_{\rm env}$, as expected.
This confirms that the sources not associated with 24~$\mu$m emission are indeed
in an earlier evolutionary phase than those associated. In fact, almost all the
24~$\mu$m-dark  sources occupy a lower part of the accretion phase of the 
Molinari et al.\ evolutionary tracks, while the 24~$\mu$m-bright ones are located
closer to the ZAMS, as indicated by the end of the ascending tracks.

\subsection{Embedded population in the G29-SFR cloud}

\begin{figure}
\centerline{\includegraphics[angle=0,width=8cm]{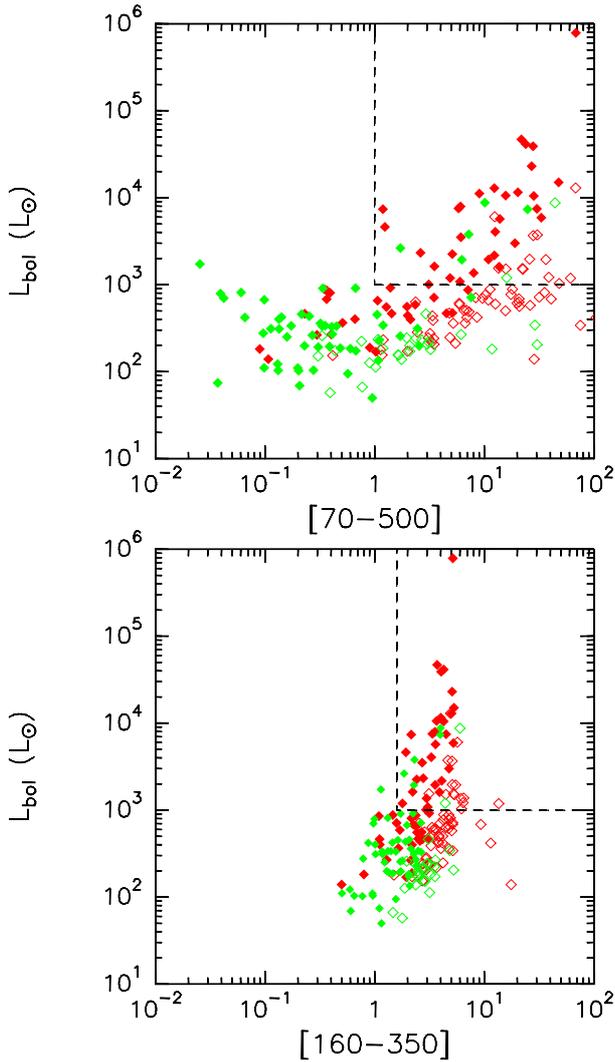}}
\caption{Luminosity--color plots for the \hi\ 24~$\mu$m-bright 
({\it red diamonds}) and 24~$\mu$m-dark ({\it green diamonds}) sources 
in the G29-SFR cloud. The empty red and green diamonds indicate sources with 
$M_{\rm env}< 100\, M_\odot$.
Dashed lines indicate the threshold of the area defined by Faimali et
al.~(\cite{faimali12}) for selecting embedded massive YSOs, at a luminosity
$>10^3\,L_\odot$.}
\label{l-color}
\end{figure}

As discussed in the previous section, most of the sources in the G29-SFR cloud 
seem to be in the main accretion pre-main sequence phase or early ZAMS phase
(Fig.~\ref{tracks}). This seems to indicate that the population in the G29-SFR
cloud, mostly massive sources, should be highly embedded. In a recent work,
Faimali et al.~(\cite{faimali12}) analyze Hi-GAL data on another massive
star-forming region G305 and propose a far-IR color criterion to select massive
embedded sources. According to these authors, the [70--500] and the [160--350]
colors should be most sensitive to the embedded population. Based on the fact
that the embedded massive protostars in G305, associated with typical signposts
of massive star formation such as free-free emission, water and/or methanol
masers, and 24~$\mu$m emission, are confined to an area of $L_{\rm bol}$-color plots,
these authors propose that embedded massive star-forming sources, both
prestellar and protostellar, should have [70--500]\,$\geq1$  and [160--350]\,$\geq1.6$
for $L_{\rm bol}>10^3\,L_\odot$. To  check whether these selection criteria for embedded massive
sources are valid for our sources, we plot the luminosity versus color in
Fig.~\ref{l-color}. The distribution of sources is very similar to that found by
Faimali et al.~(\cite{faimali12}) for the sources in G305. In the G29-SFR cloud,
we found 46 24~$\mu$m-bright and 7 24~$\mu$m-dark sources that satisfy the
criterion for embedded massive star candidates, a number similar to that found
by Faimali et al.~(\cite{faimali12}) in G305. This would indicate that only 
$\sim$27\% of the population in the G29-SFR cloud would be embedded massive star
candidates. However, as previously mentioned, most of the sources in the G29-SFR
cloud seem to be pre-main sequence sources in the main accretion phase or early
ZAMS phase, and therefore, embedded. One possible explanation for this
discrepancy could be that most of these sources in the G29-SFR cloud have
$L_{\rm bol}<$10$^3\,L_\odot$, and therefore lie by definition outside the selection
criterion area. However, by doing this, the selection criterion would miss those
young massive embedded protostars in a very early evolutionary phase that have
not yet reached their final luminosity (see Fig.~\ref{tracks}).  

The second problem with the Faimali et al.~(\cite{faimali12}) selection
criterion is that, as shown in Fig.~\ref{l-color}, there are a few sources that 
are clearly not massive ($M_{\rm env}< 100\, M_\odot$) and have $L_{\rm
bol}>$10$^3\,L_\odot$ (see Fig.~\ref{l-color}) that would fall inside the
massive embedded population area. If we lower the limit to  $M_{\rm env}< 50 \,
M_\odot$, there are still 8 24~$\mu$m-bright sources that would satisfy the
criterion. Therefore, all this suggests that the  far-IR color  selection
criterion for embedded massive YSOs of Faimali et al.~(\cite{faimali12}) cannot
be applied in all the massive star forming regions.

\subsection{The star formation efficiency and rate}

Observations of OB associations and Giant Molecular Clouds indicate that the
overall star formation efficiency, SFE=$M_{\rm stars}/(M_{\rm stars}+M_{\rm
cloud})$, is very low, $\sim$3--4\% (Evans \& Lada~\cite{evans91};
Lada~\cite{lada99}). To estimate the SFE in the G29-SFR cloud, we first need the
total gass mass of the cloud, $M_{\rm cloud}$, and the mass of the stars,
$M_{\rm stars}$. The former can be estimated from the \hi\ data, while $M_{\rm
stars}$ can be calculated by assuming that the emission in the G29-SFR cloud is
consistent with that of a stellar cluster. To check this, we calculated the Lyman
continuum, $N_{\rm Ly}$, of the cloud by measuring the radio flux at 20~cm, and
compared this value with the bolometric luminosity, $L_{\rm bol}$, of the cloud.
$L_{\rm bol}$ was calculated integrating the \hi\ emission, inside the same area used
to estimate the centimeter flux, in the 5 \hi\ bands and fitting the SED with a
modified blackbody. The total radio flux at centimeter wavelengths is 30.6~Jy,
which corresponds to $N_{\rm Ly}$=$1.08\times10^{50}$~s$^{-1}$. The total
$L_{\rm bol}$ is $2.2\times10^{6}\,L_\odot$. For comparison, the sum of $L_{\rm
bol}$ of all the sources that fall inside the area used to estimate the radio
flux at 20~cm is $1.1\times10^{6}\,L_\odot$. These values are consistent with
the expected $N_{\rm Ly}$ and $L_{\rm bol}$ of a stellar cluster according to
the simulations of a large collection ($10^6$) of clusters with sizes ranging
from 5 to 500000 stars each (L.\ Testi, private communication; see
S\'anchez-Monge et al.~\cite{sanchez12} for a description of the cluster
generation). For each cluster simulated, the total mass, bolometric luminosity,
maximum stellar mass and integrated Lyman continuum are computed. For a
bolometric luminosity of $2.2\times10^{6}\,L_\odot$, 90\% of the simulated
clusters have a total stellar mass $M_{\rm stars}$ between 600 and
4170~$M_\odot$. The total gas mass of the cloud, estimated by fitting a
modified  blackbody to the integrated emission of the cloud, inside the same
area used to estimate the radio flux at 20~cm, at the Herschel wavelengths, is
$8\times10^{4}\,M_\odot$.  Therefore, the overall SFE of the G29-SFR cloud 
ranges from 0.7 to 5\%, as low as that estimated in other molecular clouds
(Evans \& Lada~\cite{evans91}). For comparison, the sum of the masses of all the
sources that fall inside the area used to estimate the centimeter flux is
slightly smaller $3\times10^{4}\,M_\odot$, and the SFE slightly higher, from 2 to
12\%.

\begin{figure*}
\centerline{\includegraphics[angle=0,width=14cm]{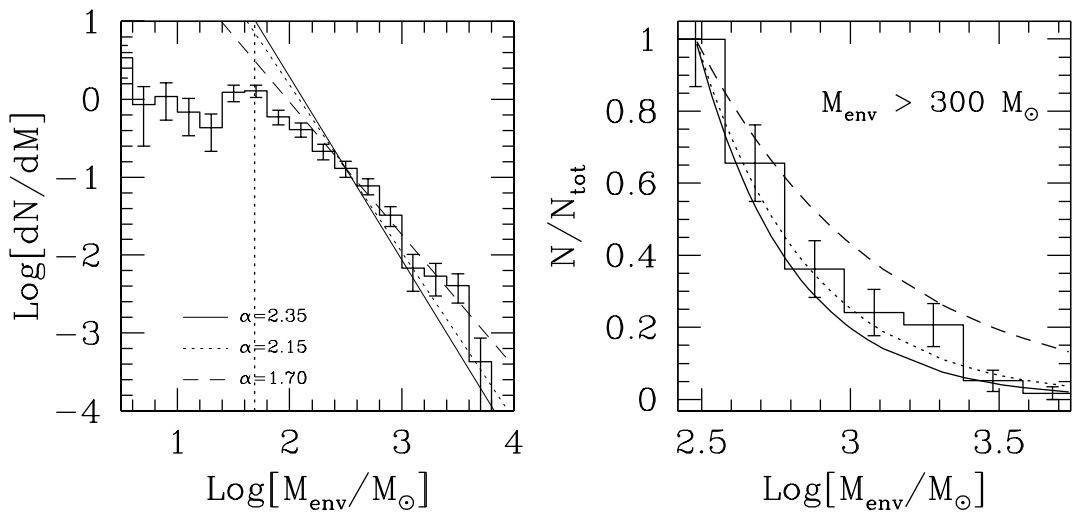}}
\caption{{\it Left panel}: the mass spectrum of the sources in the G29-SFR cloud. The solid
line represents the Salpeter IMF, d$N$/d$M \propto M_{\rm env}^{-2.35}$, the dotted line
is a $-2.15$ power law, and the dashed line is a $-1.7$ power law. The vertical dotted line indicates the completeness limit
of 49~$M_\odot$ for a temperature of 20~K. {\it Right panel}: the normalized cumulative mass distribution
of sources with masses above 300~$M_\odot$ (well above the completeness limit:
see $\S$~\ref{ms}). The solid, dotted, and dashed lines are the same as in the left panel.}
\label{imf}
\end{figure*}

The star formation rate of the cloud can be estimated as SFR=($M_{\rm
cloud}\times$\,SFE)/$t$, where $t$ is the star formation timescale needed for
the protostars to reach the ZAMS. To compare our study of the G29-SFR cloud 
with that of Faimali et al.~(\cite{faimali12}), we assume the same timescale of
0.5~Myr used by these authors, which is based on a steady-state star formation
model (Offner \& McKee~\cite{offner11}). The SFR obtained for the G29-SFR cloud
ranges from 0.001 to 0.008~$M_\odot$\,yr$^{-1}$. These values are smaller
than those of 0.01--0.02~$M_\odot$\,yr$^{-1}$ estimated by Faimali et
al.~(\cite{faimali12}) for the G305 cloud, but consistent with the values of
$\sim$0.0002--0.001~$M_\odot$\,yr$^{-1}$ estimated by Veneziani et
al.~(\cite{veneziani12}) for the whole $l=30\degr$ SDP field, and with the SFRs
of $\sim$0.0005 to $\sim$0.008~$M_\odot$\,yr$^{-1}$ estimated for Galactic \HII
regions by Chomiuk \& Povich~(\cite{chomiuk11}). The fact that the SFR of the
Milky Way is of about 2~$M_\odot$\,yr$^{-1}$ (Chomiuk \&
Povich~\cite{chomiuk11}), indicates that hundreds to a few thousands of
molecular clouds similar to the G29-SFR cloud are needed to account for the 
Galactic star formation rate.

\subsection{The clump mass function}
\label{ms}

Figure~\ref{imf} shows the mass spectrum of the sources in the G29-SFR cloud.
Olmi et al.~(\cite{olmi13}) have analyzed the whole $l=30\degr$ SDP field and
estimated a statistical mass completeness limit, from the 160~$\mu$m maps at the
80\% confidence level, of 73~$M_\odot$ for a temperature of 20~K, a dust mass
absorption coefficient $\kappa_{0} = 11$\,cm$^2$\,g$^{-1}$, evaluated at
$\nu_0=c/250\,\mu$m, and a gas-to-dust ratio of 100 (Martin et
al.~\cite{martin12}), a dust emissivity index of 2, and a median distance for
the whole field of 7.6~kpc. Assuming a distance of 6.2~Kpc for the G29-SFR
cloud, the mass completeness limit is of $\sim$49~$M_\odot$. 

If the source mass distribution can be represented by a power law of the type
$dN/dM \propto M_{\rm env}^{-\alpha}$, then the histogram of the mass spectrum
can be fitted with a straight line of slope $-\alpha$. The solid line in the
figures corresponds to $\alpha=2.35$, i.e., the Salpeter~(\cite{salpeter55})
Initial Mass Function (IMF), and the dashed line to $\alpha=1.70$, corresponding
to the mass function of molecular clouds derived from gas, mainly CO,
observations (e.g.\ Kramer et al.~\cite{kramer98}). The dotted line corresponds
to the best-fit power-law index of $\alpha=2.15\pm0.30$ obtained with a
procedure that implements both the discrete and continuous maximum likelihood
estimator for fitting the power-law distribution to data, along with a
goodness-of-fit based approach to estimating the lower cutoff of the data (see
Clauset et al.~\cite{clauset09} and Olmi et al.~\cite{olmi13} for a detailed
description of this method). This lower cutoff will be indicated here as
$M_{\rm inf}$, which will thus represent the value below which the behavior of
the distribution departs from a power-law. Following Clauset et
al.~(\cite{clauset09}), we have chosen the value of $M_{\rm inf}$ that makes the
probability distributions of the measured data and the best-fit power-law model
as similar as possible above $M_{\rm inf}$. In order to quantify the difference
between these probability distributions, the Kolmogorov-Smirnov statistics
is used. The value of $M_{\rm inf}$  for the sources in the G29-SFR cloud  is
$\sim$$300\pm130~M_\odot$. This is well above the mass completeness
limit. The right panel shows the normalized cumulative mass distribution of the
58 sources with masses above $M_{\rm inf}$.

The best-fit power-law index $\alpha$ of 2.15 obtained for the G29-SFR cloud is
the same obtained by Olmi et al.~(\cite{olmi13}) for the whole $l=30\degr$ SDP
field. $M_{\rm inf}$  is consistent within the errors with the value of 
$200\pm79~M_\odot$ obtained for the whole field. The power-law index is also
consistent with the value of 2.20 obtained by the same authors for $l=59\degr$,
the second SDP field. $M_{\rm inf}$ for this field,   $7.3\pm2.2~M_\odot$, is
much lower than the value of $\sim$$300~M_\odot$ estimated for the G29-SFR cloud,
but this is not surprising taking into account that the $l=59\degr$ region
contains mostly low- to intermediate-mass sources (the median mass for this field
is of about 2.1~$M_\odot$: Olmi et al.~\cite{olmi13}). These values of the 
power-law index $\alpha$ agree with the typical values found by Swift \& Beaumont
(\cite{swift10}), for CMFs of both low- and high-mass star-forming regions. This
suggests that from the shape of the CMF it is not possible to foresee a different
evolution towards the IMF for high- and low-mass star-forming clumps (Olmi et
al.~\cite{olmi13}). 

The value of $\alpha=2.15\pm0.30$ is also consistent within
the errors with the value of 2.35 of the stellar IMF 
(Salpeter~\cite{salpeter55}). The observational similarity between the CMF and
the IMF, first noted by Motte et al.~(\cite{motte98}) for the low-mass
star-forming region $\rho$ Ophiuchi, has been since then observed in many other
low-mass star-forming regions (e.g.\ Simpson et al.\cite{simpson08} and
references therein). This similar behavior  has inspired the idea that
gravitational fragmentation plays a key role in determining the final mass of the
stars, that is, the IMF, in clustered regions (Motte et al.~\cite{motte98}). 
That the CMF
of high-mass star-forming regions mimics the stellar IMF (this work; Beltr\'an et
al.~\cite{beltran06}) seems to suggest that also in this case, the fragmentation of massive clumps
may determine the IMF and the masses of the final stars. In other words, the
processes that determine the clump mass spectrum might be self-similar across a
broad range of clump and parent cloud masses.

\section{Conclusions}

We have conducted a far-infrared (FIR) study of the G29-SFR cloud using the \hi\
data at 70, 160, 250, 350, and 500~$\mu$m aimed at identifying the sources associated with
this high-mass star-forming region and estimate their physical properties. 

A total of 198 sources have been detected in all 5 \hi\ bands. The
mean and median values of their physical properties are 0.36 and 0.36~pc for the
radius, 379 and 115~$M_\odot$ for the mass, 0.24 and 0.06~g\,cm$^{-2}$ for the
surface density, 29 and 25~K for the temperature, $6.2\times10^3$ and
470~$L_\odot$ for the luminosity, and 23 and 5~$L_\odot/M_\odot$ for the
luminosity-to-mass ratio. 

The G29-SFR cloud is associated with 10 NVSS sources and with extended centimeter
continuum emission well correlated with the 70~$\mu$m emission. This suggests that the
cloud would contain a group of \HII regions that are ionizing and disrupting
the cloud. Assuming that the centimeter continuum emission comes from
homogeneous optically thin regions, we estimated that most of the NVSS sources
would be early B or late O types. The cloud would also contain 3 sources, with one of
them being that associated with the G29-UC region, with spectral types O5--O6.5. The study of the
distribution of masses, surface densities, luminosities, temperatures, and
luminosity-to-mass ratios of the \hi\ sources as a function of the distance to
the NVSS sources indicates that the most massive and luminous sources in the
cloud are located close to the G29-UC region. This could suggest that there is a
privileged area for massive star formation towards the center of the G29-SFR
cloud.

There are 117 \hi\ sources associated with 24~$\mu$m emission, called
24~$\mu$m-bright,  and 87 sources not associated, called 24~$\mu$m-dark. Both
groups are uniformly distributed over the cloud.  The radius of 24~$\mu$m-dark
and 24~$\mu$m-bright sources is similar, the temperature and luminosity are smaller
for the 24~$\mu$m-dark than for the 24~$\mu$m-bright objects, and the mass and
surface density are higher. The luminosity-to-mass ratio is  $\sim$5--6 times
lower for 24~$\mu$m-dark sources. The 24~$\mu$m-dark  and 24~$\mu$m-bright sources
occupy different regions of the $L_{\rm bol}$--$M_{\rm env}$  and $L_{\rm
bol}/M_{\rm env}$--$M_{\rm env}$ diagrams, with the 24~$\mu$m-dark sources having
lower $L_{\rm bol}$ and $L_{\rm bol}/M_{\rm env}$ for similar $M_{\rm env}$, as
expected. All this suggests that  the sources not associated with
24~$\mu$m emission are in an earlier evolutionary phase than those associated.
This is supported by the fact that the [70--160] color of 24~$\mu$m-dark sources
is clearly smaller than that of the 24~$\mu$m-bright ones.

Almost all the \hi\ sources in the G29-SFR cloud have masses well above the
Jeans mass and would be gravitationally supercritical if only supported by
thermal pressure. However, only $\sim$6\% of the sources have masses above
the virial mass, which confirms that an additional supporting agent, such as
turbulence, might be acting against gravity in these sources. The percentage of
sources with masses larger than the virial mass is clearly higher for those 
located at $\lesssim 4'$ of the G29-UC region. This suggests that the sources
that should be undergoing collapse and forming stars are preferentially
concentrated towards the dominant source in the cloud.

The overall SFE of the G29-SFR cloud ranges from 0.7 to 5\%, and it is as low as
that estimated in other molecular clouds. The SFR  ranges from 0.001 to
0.008~$M_\odot$\,yr$^{-1}$ and is consistent with the values estimated for
Galactic \HII regions. To account for the SFR of 2~$M_\odot$\,yr$^{-1}$ of the
Milky Way, hundreds to a few thousands of molecular clouds similar to the
G29-SFR cloud would be needed. 

The mass spectrum of the \hi\ sources with masses above $300~M_\odot$, well above
the completeness limit, can be well-fitted with a power law of slope
$\alpha=2.15\pm0.30$, consistent with the values obtained by Olmi et
al.~(\cite{olmi13}) for the whole $l=30\degr$, associated with high-mass star
formation, and $l=59\degr$, associated with low- to intermediate-mass star
formation, \hi\ SDP fields. The observational similarity of the CMF for low- and
high-mass star-forming regions suggests that from the CMF itself is not possible
to predict a different evolution of the clumps towards the IMF. The fact that the
CMF of the G29-SFR cloud mimics, within the errors, the stellar IMF suggests a
self-similar process which determines the shape of the mass spectrum over a broad
range of masses, from stellar to cluster size scales.

\begin{acknowledgements} 

It is a pleasure to thank Annie Zavagno for critically reading the manuscript.
Hi-GAL data processing and analysis has been possible thanks to the Italian Space
Agency support via contract I/038/080/0. SPIRE has been developed by a consortium
of institutes led by Cardiff Univ. (UK) and including: Univ. Lethbridge (Canada);
NAOC (China); CEA, LAM (France); IFSI, Univ. Padua (Italy); IAC (Spain);
Stockholm Observatory (Sweden); Imperial College London, RAL, UCL-MSSL, UKATC,
Univ. Sussex (UK); and Caltech, JPL, NHSC, Univ. Colorado (USA). This development
has been supported by national funding agencies: CSA (Canada); NAOC (China); CEA,
CNES, CNRS (France); ASI (Italy); MCINN (Spain); SNSB (Sweden); STFC, UKSA (UK);
and NASA (USA). PACS has been developed by a consortium of institutes led by MPE
(Germany) and including UVIE (Austria); KU Leuven, CSL, IMEC (Belgium); CEA, LAM
(France); MPIA (Germany); INAF-IFSI/OAA/OAP/OAT, LENS, SISSA (Italy); IAC
(Spain). This development has been supported by the funding agencies BMVIT
(Austria), ESAPRODEX (Belgium), CEA/CNES (France), DLR (Germany), ASI/INAF
(Italy), and CICYT/MCYT (Spain). This publication makes use of data products from
the Wide-field Infrared Survey Explorer, which is a joint project of the
University of California, Los Angeles, and the Jet Propulsion
Laboratory/California Institute of Technology, funded by the National Aeronautics
and Space Administration. This research made use of data products from the
Midcourse Space Experiment. Processing of the data was funded by the Ballistic
Missile Defense Organization with additional support from NASA Office of Space
Science. This research has also made use of the NASA/IPAC Infrared Science
Archive, which is operated by the Jet Propulsion Laboratory, California Institute
of Technology, under contract with the National Aeronautics and Space
Administration.  \end{acknowledgements}

\clearpage

\addtocounter{table}{-6}
\renewcommand{\thefootnote}{\alph{footnote}}
\onecolumn
\begin{scriptsize}
\begin{longtable}{lcccccccccc}
\caption[]{Position, \hi\ fluxes, and MIPSGAL 24~$\mu$m association 
for the sources detected by {\it Herschel} towards the G29.96$-0.02$ cloud} \\

\hline 
&\multicolumn{1}{c}{$\alpha({\rm J2000})$} &
\multicolumn{1}{c}{$\delta({\rm J2000})$} &
\multicolumn{1}{c}{$l$} &
\multicolumn{1}{c}{$b$} &
\multicolumn{1}{c}{$S_{70 \mu m}$} &
\multicolumn{1}{c}{$S_{160 \mu m}$} &
\multicolumn{1}{c}{$S_{250 \mu m}$} &
\multicolumn{1}{c}{$S_{350 \mu m}$} &
\multicolumn{1}{c}{$S_{500 \mu m}$} & \\
\multicolumn{1}{c}{\# Id.} &
\multicolumn{1}{c}{($^{\rm h}$ $^{\rm m}$ $^{\rm s}$)}&
\multicolumn{1}{c}{($\degr$ $\arcmin$ $\arcsec$)} &
\multicolumn{1}{c}{($\degr$)}&
\multicolumn{1}{c}{($\degr$)}&
\multicolumn{1}{c}{(Jy)} &
\multicolumn{1}{c}{(Jy)} &
\multicolumn{1}{c}{(Jy)} &
\multicolumn{1}{c}{(Jy)} &
\multicolumn{1}{c}{(Jy)} & 
\multicolumn{1}{c}{24~$\mu$m MIPS} \\
\hline \\[-1.8ex]
\endfirsthead
\multicolumn{11}{c}{{\tablename} \thetable{} -- Continued} \\[0.5ex]
\hline \\[-2ex]
&\multicolumn{1}{c}{$\alpha({\rm J2000})$} &
\multicolumn{1}{c}{$\delta({\rm J2000})$} &
\multicolumn{1}{c}{$l$} &
\multicolumn{1}{c}{$b$} &
\multicolumn{1}{c}{$S_{70 \mu m}$} &
\multicolumn{1}{c}{$S_{160 \mu m}$} &
\multicolumn{1}{c}{$S_{250 \mu m}$} &
\multicolumn{1}{c}{$S_{350 \mu m}$} &
\multicolumn{1}{c}{$S_{500 \mu m}$} & \\
\multicolumn{1}{c}{\# Id.} &
\multicolumn{1}{c}{($^{\rm h}$ $^{\rm m}$ $^{\rm s}$)}&
\multicolumn{1}{c}{($\degr$ $\arcmin$ $\arcsec$)} &
\multicolumn{1}{c}{($\degr$)}&
\multicolumn{1}{c}{($\degr$)}&
\multicolumn{1}{c}{(Jy)} &
\multicolumn{1}{c}{(Jy)} &
\multicolumn{1}{c}{(Jy)} &
\multicolumn{1}{c}{(Jy)} &
\multicolumn{1}{c}{(Jy)} & 
\multicolumn{1}{c}{24~$\mu$m MIPS} \\
\hline \\[-1.8ex]
\endhead
  1 &18 46 06.05 &$-$2 41 18.3 &29.93 &$-$0.04 &    55$\pm$9   &  58$\pm$8   &  38$\pm$4    & 25$\pm$3	&   7.5$\pm$1.1 &N \\
  2 &18 45 51.92 &$-$2 42 23.8 &29.89 &$+$0.00 &    99$\pm$12  & 156$\pm$17  &  95$\pm$11   & 46$\pm$6	& 17$\pm$2 &Y \\
  4 &18 46 11.67 &$-$2 38 37.7 &29.98 &$-$0.04 &   0.59$\pm$0.11   &  24$\pm$5   &  36$\pm$8  & 24$\pm$5	& 14$\pm$3 &N \\
  5 &18 45 59.57 &$-$2 43 10.4 &29.89 &$-$0.03 &   0.70$\pm$0.13   &  19$\pm$5   &  14$\pm$3  &  9.8$\pm$2.2	&  2.8$\pm$0.7 &N \\
  6 &18 45 57.49 &$-$2 44 04.1 &29.87 &$-$0.03 &   0.62$\pm$0.11   &  15$\pm$4   &  16$\pm$3  & 10$\pm$2      &  4.3$\pm$1.0 &N \\
  7 &18 46 05.63 &$-$2 44 32.8 &29.88 &$-$0.06 &   0.72$\pm$0.13   &  14$\pm$4   &  12$\pm$3  &  6.6$\pm$1.7      &  2.2$\pm$0.6 &N \\
  8 &18 46 06.38 &$-$2 44 45.9 &29.88 &$-$0.07 &   0.83$\pm$0.15   &  12$\pm$1   &   8.2$\pm$0.9    &  4.8$\pm$0.7	&  2.0$\pm$0.3 &N \\
  9 &18 46 25.08 &$-$2 37 52.2 &30.02 &$-$0.08 &   0.34$\pm$0.06   &   4.1$\pm$0.6   &   6.2$\pm$0.8    &  4.3$\pm$0.6	&  1.7$\pm$0.3 &N \\
 11 &18 46 26.19 &$-$2 37 03.1 &30.03 &$-$0.08 &   1.9$\pm$0.3   &   6.5$\pm$0.8   &   4.7$\pm$0.6    &  2.0$\pm$0.3	&  0.7$\pm$0.1 &N \\
 12 &18 45 46.50 &$-$2 33 14.1 &30.01 &$+$0.09 &   0.36$\pm$0.07   &   3.8$\pm$1.2   &   9.5$\pm$1.2    &  6.4$\pm$0.9	&  2.8$\pm$0.4 &N \\
 13 &18 45 48.53 &$-$2 37 40.9 &29.95 &$+$0.05 &   0.48$\pm$0.09   &   8.9$\pm$1.3   &  12$\pm$2    &  7.2$\pm$1.1	&  3.0$\pm$0.5 &N \\
 14 &18 46 07.43 &$-$2 34 06.0 &30.04 &$+$0.01 &   3.8$\pm$0.43   &  10$\pm$1   &   8.6$\pm$1.0    &  4.7$\pm$0.6	&  1.8$\pm$0.2 &Y \\
 16 &18 46 41.10 &$-$2 36 28.9 &30.07 &$-$0.13 &   0.88$\pm$0.17   &   5.5$\pm$0.6   &   6.1$\pm$0.7    &  3.8$\pm$0.5	&  1.8$\pm$0.3 &N \\
 17 &18 46 40.03 &$-$2 41 49.3 &29.99 &$-$0.17 &   2.7$\pm$0.3   &   8.9$\pm$1.0   &  11$\pm$1    &  7.9$\pm$1.0	&  4.0$\pm$0.6 &Y \\
 18 &18 46 41.04 &$-$2 37 05.8 &30.06 &$-$0.14 &   0.05$\pm$0.05   &   2.8$\pm$1.8   &   3.8$\pm$2.3    & 2.6$\pm$1.6 &  1.4$\pm$0.8 &N \\
 19 &18 46 42.12 &$-$2 37 28.4 &30.06 &$-$0.14 &   0.71$\pm$0.13   &   6.1$\pm$0.8   &   5.7$\pm$0.7    &  3.0$\pm$0.4	&  1.2$\pm$0.2 &N \\
 20 &18 46 40.78 &$-$2 39 44.2 &30.02 &$-$0.16 &   0.42$\pm$0.08   &   3.5$\pm$0.4   &   6.4$\pm$0.7    &  4.5$\pm$0.6	&  2.1$\pm$0.3 &N \\
 21 &18 46 39.64 &$-$2 35 14.3 &30.08 &$-$0.12 &   0.85$\pm$0.15   &   9.6$\pm$1.1   &   6.0$\pm$0.7    &  2.7$\pm$0.4	&  1.1$\pm$0.2 &N \\
 22 &18 46 42.42 &$-$2 40 07.0 &30.02 &$-$0.16 &   1.8$\pm$0.25   &   7.7$\pm$0.9   &   5.7$\pm$0.6    &  2.5$\pm$0.3	&  0.9$\pm$0.1 &N \\
 24 &18 46 16.79 &$-$2 34 56.4 &30.05 &$-$0.03 &   0.85$\pm$0.11   &   4.8$\pm$0.6   &   5.3$\pm$0.6    &  3.2$\pm$0.4	&  0.7$\pm$0.1 &Y \\
 25 &18 46 31.69 &$-$2 37 13.6 &30.04 &$-$0.10 &   0.64$\pm$0.12   &   6.9$\pm$0.8   &   7.1$\pm$0.8    &  4.0$\pm$0.5	&  1.6$\pm$0.2 &N \\
 26 &18 46 15.19 &$-$2 34 27.6 &30.05 &$-$0.02 &   0.48$\pm$0.09   &   3.6$\pm$0.4   &   6.1$\pm$0.7    &  3.7$\pm$0.5	&  1.7$\pm$0.2 &N \\
 27 &18 46 40.23 &$-$2 34 35.7 &30.10 &$-$0.11 &   1.4$\pm$0.2   &   7.2$\pm$1.0   &   5.9$\pm$0.8    &  3.0$\pm$0.4	&  1.3$\pm$0.2 &N \\
 31 &18 46 36.02 &$-$2 42 40.3 &29.97 &$-$0.16 &   0.38$\pm$0.07   &   3.5$\pm$0.4   &   6.9$\pm$0.8    &  5.3$\pm$0.7	&  2.9$\pm$0.4 &N \\
 33 &18 45 43.88 &$-$2 37 55.5 &29.94 &$+$0.07 &   2.3$\pm$0.3   &   7.3$\pm$0.821   &   5.6$\pm$0.6    &  2.5$\pm$0.3	&  0.7$\pm$0.1 &Y \\
 34 &18 46 27.23 &$-$2 34 06.7 &30.08 &$-$0.06 &   0.71$\pm$0.13   &   5.8$\pm$0.7   &   5.3$\pm$0.6    &  2.8$\pm$0.4	&  1.0$\pm$0.1 &N \\
 35 &18 45 49.20 &$-$2 35 18.6 &29.99 &$+$0.07 &   1.8$\pm$0.2   &   6.7$\pm$0.7   &   5.1$\pm$0.6    &  2.2$\pm$0.3	&  0.8$\pm$0.1 &N \\
 37 &18 45 47.36 &$-$2 36 05.6 &29.97 &$+$0.07 &   1.1$\pm$0.2   &   6.3$\pm$1.1   &   5.0$\pm$0.9    &  2.9$\pm$0.5	&  1.2$\pm$0.2 &Y \\
 38 &18 46 52.32 &$-$2 39 16.4 &30.05 &$-$0.20 &   2.0$\pm$0.3   &   5.2$\pm$0.6   &   3.8$\pm$0.4    &  1.8$\pm$0.2	&  0.6$\pm$0.1 &N \\
 40 &18 45 48.46 &$-$2 34 47.7 &29.99 &$+$0.08 &   2.2$\pm$0.4   &   5.1$\pm$0.9   &   3.8$\pm$0.6    &  1.9$\pm$0.3	&  0.7$\pm$0.1 &N \\
 42 &18 46 22.46 &$-$2 34 06.8 &30.07 &$-$0.05 &   2.1$\pm$0.3   &   5.5$\pm$0.6   &   3.9$\pm$0.5    &  1.8$\pm$0.2	&  0.6$\pm$0.1 &Y \\
 44 &18 46 19.25 &$-$2 46 40.4 &29.88 &$-$0.13 &   3.4$\pm$0.5   &   7.3$\pm$0.9   &   5.0$\pm$0.6    &  3.2$\pm$0.4	&  1.4$\pm$0.2 &N \\
 45 &18 45 59.56 &$-$2 49 42.1 &29.79 &$-$0.08 &   0.54$\pm$0.10   &   6.1$\pm$0.7   &   7.7$\pm$0.9    &  4.5$\pm$0.6	&  1.8$\pm$0.2 &N \\
 48 &18 46 16.81 &$-$2 33 47.2 &30.06 &$-$0.02 &   3.3$\pm$0.4   &   5.8$\pm$0.6   &   3.5$\pm$0.4    &  1.6$\pm$0.2	&  0.5$\pm$0.1 &N \\
 49 &18 46 55.28 &$-$2 36 33.1 &30.10 &$-$0.19 &   1.8$\pm$0.3   &   5.1$\pm$0.6   &   4.1$\pm$0.5    &  2.0$\pm$0.3	&  0.8$\pm$0.1 &N \\
 54 &18 46 28.85 &$-$2 31 14.1 &30.12 &$-$0.05 &   2.0$\pm$0.2   &   3.5$\pm$0.4   &   2.7$\pm$0.3    &  1.1$\pm$0.1	&  0.2$\pm$0.03 &N \\
 55 &18 46 58.48 &$-$2 35 22.8 &30.12 &$-$0.19 &   0.67$\pm$0.12   &   4.1$\pm$0.5   &   4.4$\pm$0.5    &  2.6$\pm$0.3	&  1.2$\pm$0.2 &N \\
 61 &18 46 49.82 &$-$2 33 42.6 &30.13 &$-$0.14 &   0.45$\pm$0.11   &   6.7$\pm$0.7   &   5.1$\pm$0.6    &  2.6$\pm$0.3	&  1.0$\pm$0.1 &N \\
 62 &18 46 28.18 &$-$2 50 00.2 &29.84 &$-$0.19 &   1.0$\pm$0.1   &   4.0$\pm$0.4   &   3.5$\pm$0.4    &  1.9$\pm$0.2	&  0.9$\pm$0.1 &N \\
 64 &18 46 37.41 &$-$2 45 28.8 &29.93 &$-$0.19 &   1.1$\pm$0.1   &   5.0$\pm$0.6   &   4.6$\pm$0.5    &  2.6$\pm$0.3	&  1.0$\pm$0.1 &Y \\
 65 &18 45 56.00 &$-$2 49 46.6 &29.79 &$-$0.07 &   2.3$\pm$0.3   &   6.4$\pm$0.7   &   4.4$\pm$0.5    &  2.0$\pm$0.3	&  0.7$\pm$0.1 &Y \\
 66 &18 46 28.51 &$-$2 49 15.7 &29.86 &$-$0.18 &   1.4$\pm$0.2   &   3.9$\pm$0.5   &   3.3$\pm$0.4    &  1.8$\pm$0.2	&  0.8$\pm$0.1 &N \\
 69 &18 46 17.94 &$-$2 51 46.3 &29.80 &$-$0.16 &   6.8$\pm$0.8   &   9.6$\pm$1.1   &   5.5$\pm$0.6    &  2.5$\pm$0.3	&  0.9$\pm$0.1 &Y \\
 70 &18 46 27.79 &$-$2 51 34.9 &29.82 &$-$0.20 &   0.32$\pm$0.06   &   3.1$\pm$0.4   &   7.7$\pm$0.9    &  6.2$\pm$0.8	&  3.3$\pm$0.5 &N \\
 73 &18 46 27.25 &$-$2 50 50.9 &29.83 &$-$0.19 &   0.42$\pm$0.08   &   2.1$\pm$0.3   &   4.7$\pm$0.5    &  3.5$\pm$0.5	&  2.0$\pm$0.3 &N \\
 74 &18 45 53.61 &$-$2 49 27.9 &29.79 &$-$0.06 &   0.28$\pm$0.19   &   7.5$\pm$0.9   &   5.5$\pm$0.6    &  2.6$\pm$0.3	&  0.9$\pm$0.1 &N \\
 75 &18 46 29.85 &$-$2 45 18.8 &29.92 &$-$0.16 &   0.39$\pm$0.13   &   4.3$\pm$0.5   &   3.1$\pm$0.3    &  1.4$\pm$0.2	&  0.4$\pm$0.1 &N \\
 77 &18 46 23.22 &$-$2 50 44.3 &29.82 &$-$0.18 &   1.9$\pm$0.3   &   4.0$\pm$0.5   &   3.3$\pm$0.4    &  1.6$\pm$0.2	&  0.6$\pm$0.1 &N \\
 79 &18 46 12.13 &$-$2 51 47.1 &29.79 &$-$0.14 &   1.7$\pm$0.2   &   8.1$\pm$0.9   &   5.9$\pm$0.7    &  2.7$\pm$0.4	&  0.7$\pm$0.1 &Y \\
 81 &18 46 40.58 &$-$2 45 45.8 &29.93 &$-$0.20 &   2.8$\pm$0.4   &   5.5$\pm$0.7   &   3.8$\pm$0.5    &  2.0$\pm$0.3	&  0.8$\pm$0.1 &Y \\
 83 &18 47 00.90 &$-$2 35 54.5 &30.12 &$-$0.20 &  0.63$\pm$0.11   &   2.8$\pm$0.3   &   3.1$\pm$0.3    &  1.9$\pm$0.3	&  0.8$\pm$0.1 &N \\
 86 &18 46 36.87 &$-$2 46 22.0 &29.91 &$-$0.19 &   0.13$\pm$0.05   &   2.0$\pm$0.4   &   2.1$\pm$0.4    &  1.1$\pm$0.2	&  0.3$\pm$0.1 &N \\
 87 &18 46 34.97 &$-$2 45 54.8 &29.92 &$-$0.18 &   0.89$\pm$0.14   &   3.8$\pm$0.5   &   3.3$\pm$0.4    &  1.6$\pm$0.2	&  0.5$\pm$0.1 &N \\
 88 &18 46 46.84 &$-$2 43 32.0 &29.98 &$-$0.21 &   0.89$\pm$0.16   &   5.8$\pm$0.6   &   4.6$\pm$0.5    &  2.3$\pm$0.3	&  0.7$\pm$0.1 &N \\
 90 &18 46 27.24 &$-$2 47 28.4 &29.88 &$-$0.16 &   1.4$\pm$0.2   &   4.3$\pm$0.6   &   3.8$\pm$0.5    &  2.0$\pm$0.3	&  0.7$\pm$0.1 &Y \\
 98 &18 45 35.30 &$-$2 38 33.8 &29.91 &$+$0.10 &   1.6$\pm$0.2   &  11$\pm$1   &  10$\pm$1    &  6.7$\pm$0.9	        &  3.4$\pm$0.5 &N \\
 99 &18 45 37.76 &$-$2 35 56.5 &29.96 &$+$0.11 &   2.7$\pm$0.3   &   3.5$\pm$0.4   &   3.2$\pm$0.4    &  2.0$\pm$0.3	&  1.1$\pm$0.1 &N \\
100 &18 46 44.22 &$-$2 43 44.2 &29.97 &$-$0.20 &   2.6$\pm$0.5   &   3.6$\pm$0.6   &   2.0$\pm$0.4    &  0.7$\pm$0.1	&  0.09$\pm$0.02 &N \\
104 &18 46 17.83 &$-$2 29 50.3 &30.12 &$+$0.00 &   0.70$\pm$0.11   &   5.3$\pm$0.6   &   3.8$\pm$0.4    &  1.9$\pm$0.2  &  0.6$\pm$0.1 &Y \\
109 &18 45 36.99 &$-$2 41 25.1 &29.87 &$+$0.07 &   0.97$\pm$0.18   &   4.6$\pm$0.6   &   3.5$\pm$0.4    &  1.6$\pm$0.2  &  0.6$\pm$0.1 &N \\
111 &18 45 35.83 &$-$2 40 00.6 &29.89 &$+$0.08 &   0.70$\pm$0.13   &   3.7$\pm$0.6   &   3.5$\pm$0.6    &  1.9$\pm$0.3        &  0.8$\pm$0.1 &N \\
113 &18 45 31.58 &$-$2 39 33.8 &29.89 &$+$0.10 &   0.81$\pm$0.15   &   2.0$\pm$0.7   &   2.4$\pm$0.8    &  1.8$\pm$0.6	&  0.8$\pm$0.3 &N \\
122 &18 46 09.87 &$-$2 41 08.1 &29.94 &$-$0.05 &  187$\pm$21  & 219$\pm$25  & 117$\pm$13   & 44$\pm$6	& 15$\pm$2 &Y \\  
123 &18 46 05.00 &$-$2 42 23.6 &29.91 &$-$0.04 &  46$\pm$8   & 210$\pm$24  & 171$\pm$19   & 97$\pm$13	& 38$\pm$5 &Y \\
124 &18 46 08.76 &$-$2 42 01.8 &29.93 &$-$0.05 &  39$\pm$12  &  77$\pm$22  &  57$\pm$16   & 28$\pm$8	&  6$\pm$2 &Y \\
125 &18 46 11.87 &$-$2 41 30.7 &29.94 &$-$0.06 & 184$\pm$27  & 162$\pm$22  &  79$\pm$11   & 40$\pm$6	& 20$\pm$3 &Y \\  
126 &18 46 12.87 &$-$2 38 58.3 &29.98 &$-$0.05 &  36$\pm$4   & 128$\pm$14  & 111$\pm$12   & 66$\pm$9	& 29$\pm$4 &Y \\
127 &18 46 00.41 &$-$2 41 14.9 &29.92 &$-$0.02 & 122$\pm$15  & 150$\pm$17  &  85$\pm$10    & 37$\pm$5	& 12$\pm$2 &N \\
129 &18 45 59.01 &$-$2 41 10.1 &29.92 &$-$0.01 &  25$\pm$4   &  63$\pm$9   &  62$\pm$9    & 34$\pm$5	& 15$\pm$2 &N \\
130 &18 46 12.92 &$-$2 39 29.6 &29.97 &$-$0.05 &   0.65$\pm$0.12   &  64$\pm$7   &  92$\pm$10   & 56$\pm$7	& 25$\pm$4 &N \\
131 &18 46 06.45 &$-$2 37 49.2 &29.98 &$-$0.01 &   0.65$\pm$0.12   &  36$\pm$2  &  39$\pm$29   & 27$\pm$20	& 11$\pm$8 &N \\
132 &18 46 10.96 &$-$2 43 28.2 &29.91 &$-$0.07 &  52$\pm$8   &  37$\pm$5   &  19$\pm$3    &  7.9$\pm$1.2	&  1.7$\pm$0.3 &Y \\
133 &18 46 13.16 &$-$2 36 35.6 &30.01 &$-$0.03 &   0.68$\pm$0.12   &  33$\pm$8   &  28$\pm$6    & 17$\pm$3.9	&  6.9$\pm$1.7 &N \\
134 &18 45 58.73 &$-$2 40 32.7 &29.93 &$-$0.01 &   0.58$\pm$0.10   &  28$\pm$3   &  42$\pm$5    & 28$\pm$3.6	& 15$\pm$2 &N \\
135 &18 46 13.04 &$-$2 43 37.9 &29.91 &$-$0.08 &  26$\pm$21  &   8.4$\pm$7   &   3.0$\pm$2.5    &  1.4$\pm$1.1	&  0.09$\pm$0.02 &Y \\
136 &18 46 13.66 &$-$2 37 29.1 &30.00 &$-$0.04 &   0.51$\pm$0.09   &  12$\pm$3   &  14$\pm$2    &  8.6$\pm$1.6	&  2.9$\pm$0.6 &N \\
137 &18 45 55.11 &$-$2 39 19.5 &29.94 &$+$0.02 &  18$\pm$2   &  61$\pm$7   &  46$\pm$5    & 22$\pm$2.8	&  6.9$\pm$1.0 &Y \\
138 &18 46 17.21 &$-$2 38 17.4 &30.00 &$-$0.06 &   7.4$\pm$1.1   &  15$\pm$2   &  12$\pm$2    &  6.0$\pm$0.9	&  0.88$\pm$0.21 &Y \\
139 &18 46 23.72 &$-$2 41 01.0 &29.97 &$-$0.10 &  23$\pm$3   &  23$\pm$3   &  15$\pm$2    &  7.2$\pm$0.9	&  1.7$\pm$0.2 &Y \\
141 &18 46 07.15 &$-$2 44 58.5 &29.88 &$-$0.07 &   5.5$\pm$2.7   &   5.5$\pm$2.6   &   2.6$\pm$1.2    &  1.1$\pm$0.6	&  0.19$\pm$0.16 &N \\
142 &18 45 52.10 &$-$2 43 46.4 &29.87 &$-$0.01 &   0.65$\pm$0.12   &  20$\pm$3   &  21$\pm$2    &  9.8$\pm$1.3	&  3.0$\pm$0.4 &N \\
143 &18 46 17.64 &$-$2 38 06.9 &30.00 &$-$0.06 &   2.7$\pm$1.8   &   1.8$\pm$1.3   &   0.89$\pm$0.79    &  0.10$\pm$0.02	&  0.09$\pm$0.02 &Y \\
144 &18 46 22.17 &$-$2 37 04.6 &30.02 &$-$0.07 &  11$\pm$2   &   9.8$\pm$2.2   &   7.2$\pm$1.8    &  3.7$\pm$1.1	&  1.4$\pm$0.5 &N \\
147 &18 46 20.96 &$-$2 38 57.5 &30.00 &$-$0.08 &  16$\pm$2   &  11$\pm$1   &   5.9$\pm$0.7    &  1.8$\pm$0.2	&  0.33$\pm$0.06 &Y \\
148 &18 45 54.33 &$-$2 38 21.9 &29.95 &$+$0.03 &   2.6$\pm$0.4   &  30$\pm$3   &  27$\pm$3    & 14$\pm$2	&  6.6$\pm$0.9 &Y \\
149 &18 45 53.99 &$-$2 38 52.9 &29.94 &$+$0.02 &   3.6$\pm$0.5   &  30$\pm$4   &  25$\pm$3    & 13$\pm$2	&  5.4$\pm$0.8 &N \\
150 &18 45 55.02 &$-$2 45 59.7 &29.84 &$-$0.03 &   1.9$\pm$0.4   &  30$\pm$4   &  31$\pm$4    & 17$\pm$2	&  5.7$\pm$0.8 &N \\
151 &18 45 54.60 &$-$2 45 42.7 &29.84 &$-$0.03 &   0.69$\pm$0.13   &  11$\pm$2   &  13$\pm$2    &  9.3$\pm$1.3	&  5.1$\pm$0.8 &N \\
152 &18 46 03.88 &$-$2 48 31.2 &29.82 &$-$0.09 &   4.2$\pm$0.8   &  32$\pm$4   &  22$\pm$2    &  10$\pm$1	&  3.0$\pm$0.4 &Y \\
153 &18 46 01.99 &$-$2 35 29.2 &30.01 &$+$0.02 &  23$\pm$3   &  23$\pm$3   &  13$\pm$1    &  5.2$\pm$0.7	&  1.5$\pm$0.2 &Y \\
155 &18 45 46.45 &$-$2 42 47.2 &29.87 &$+$0.02 &   1.4$\pm$0.7   &  11$\pm$2   &  11$\pm$2    &  9.2$\pm$2.0    &  4.2$\pm$0.9 &N \\
159 &18 45 47.89 &$-$2 44 39.4 &29.85 &$+$0.00 &   2.8$\pm$0.4   &  27$\pm$3   &  30$\pm$3    & 19$\pm$2    &  7.8$\pm$1.1 &Y \\
160 &18 45 55.86 &$-$2 37 23.3 &29.97 &$+$0.03 &  10$\pm$1   &  17$\pm$2   &  11$\pm$1    &  4.0$\pm$0.5	&  0.57$\pm$0.10 &Y \\
161 &18 45 53.42 &$-$2 45 27.1 &29.85 &$-$0.02 &   0.87$\pm$0.16   &  11$\pm$2   &   9.5$\pm$1.3    &  4.4$\pm$0.7      &  0.81$\pm$0.17 &N \\
162 &18 46 15.53 &$-$2 44 18.5 &29.90 &$-$0.10 &   4.8$\pm$1.4   &   9.8$\pm$1   &   9.6$\pm$1.2    &  5.7$\pm$0.8      &  2.8$\pm$0.4 &N \\
163 &18 46 01.89 &$-$2 47 00.7 &29.84 &$-$0.07 &   0.65$\pm$0.12   &  10$\pm$8   &  10$\pm$8    &  5.8$\pm$4.4	&  2.4$\pm$1.8 &N \\
164 &18 45 51.09 &$-$2 44 30.4 &29.86 &$-$0.01 &   0.65$\pm$0.12   &   6.1$\pm$1.5   &   7.2$\pm$1.7    &  4.7$\pm$1.2	&  2.8$\pm$0.7 &N \\
165 &18 46 08.89 &$-$2 35 16.5 &30.03 &$-$0.00 &   5.7$\pm$0.7   &  12$\pm$1  &  12$\pm$1    &  7.0$\pm$0.9	&  2.4$\pm$0.3 &Y \\
167 &18 46 11.40 &$-$2 48 24.1 &29.84 &$-$0.11 &   10$\pm$1   &  12$\pm$2   &   7.9$\pm$1.0    &  3.2$\pm$0.5	&  1.0$\pm$0.2 &Y \\
168 &18 46 01.66 &$-$2 47 49.7 &29.83 &$-$0.07 &   3.5$\pm$0.6   &  22$\pm$2   &  17$\pm$2    &  9.0$\pm$1.2	&  3.3$\pm$0.5 &Y \\
169 &18 45 44.69 &$-$2 42 24.9 &29.87 &$+$0.03 &   5.0$\pm$0.7   &   9.6$\pm$1.1   &   7.0$\pm$0.8    &  3.7$\pm$0.6	&  2.5$\pm$0.4 &Y \\
170 &18 46 29.94 &$-$2 36 27.1 &30.05 &$-$0.09 &   4.5$\pm$0.6   &  16$\pm$2   &  12$\pm$1    &  6.0$\pm$0.8	&  2.2$\pm$0.3 &Y \\
171 &18 45 59.89 &$-$2 47 25.5 &29.83 &$-$0.06 &   0.59$\pm$0.11   &  13$\pm$1   &  20$\pm$2    & 12$\pm$2	&  4.4$\pm$0.6 &N \\
172 &18 45 42.87 &$-$2 42 53.5 &29.86 &$+$0.03 &   4.3$\pm$0.6   &  21$\pm$3   &  27$\pm$3    & 19$\pm$3	& 12$\pm$2 &Y \\
175 &18 46 31.77 &$-$2 39 33.6 &30.01 &$-$0.12 &   0.56$\pm$0.10   &   7.6$\pm$0.9   &  12$\pm$1    &  9.7$\pm$1.3	&  5.8$\pm$0.8 &N \\
177 &18 46 05.18 &$-$2 30 09.6 &30.09 &$+$0.05 &  11$\pm$1       &  26$\pm$3   &  18$\pm$2    &  8.9$\pm$1.2	    &  3.6$\pm$0.5 &Y \\
178 &18 46 47.22 &$-$2 39 36.4 &30.03 &$-$0.18 &   3.7$\pm$0.5   &  17$\pm$2   &  14$\pm$2    &  6.8$\pm$0.9	&  2.9$\pm$0.4 &Y \\
179 &18 46 42.73 &$-$2 35 41.6 &30.08 &$-$0.13 &   6.1$\pm$0.7   &   6.5$\pm$0.7   &   3.6$\pm$0.4    &  1.7$\pm$0.2	&  0.71$\pm$0.11 &Y \\
180 &18 46 50.21 &$-$2 41 36.5 &30.01 &$-$0.21 &   7.6$\pm$0.8   &  14$\pm$2   &   8.5$\pm$0.9    &  3.8$\pm$0.5	&  1.3$\pm$0.2 &Y \\
182 &18 45 56.20 &$-$2 47 13.3 &29.82 &$-$0.05 &   1.6$\pm$0.4   &  27$\pm$3   &  23$\pm$3    & 12$\pm$2	&  4.2$\pm$0.6 &Y \\
184 &18 45 44.88 &$-$2 43 31.6 &29.86 &$+$0.02 &   1.6$\pm$0.5   &  14$\pm$2   &  17$\pm$2    & 12$\pm$2	&  7.1$\pm$1.1 &Y \\
185 &18 46 23.02 &$-$2 43 49.3 &29.93 &$-$0.12 &   4.3$\pm$0.6   &   7.5$\pm$0.9   &   4.8$\pm$0.5    &  1.9$\pm$0.2	&  0.64$\pm$0.09 &Y \\
186 &18 46 15.38 &$-$2 49 44.0 &29.82 &$-$0.14 &   4.0$\pm$0.6   &   8.9$\pm$1.0   &   4.6$\pm$0.5    &  1.9$\pm$0.2	&  0.77$\pm$0.11 &Y \\
187 &18 45 49.75 &$-$2 32 48.2 &30.03 &$+$0.09 &   5.2$\pm$0.6   &  19$\pm$2   &  14$\pm$2    &  5.6$\pm$0.7	&  2.2$\pm$0.3 &Y \\
188 &18 46 46.27 &$-$2 36 20.0 &30.08 &$-$0.15 &   2.1$\pm$0.3   &   6.1$\pm$0.7   &   4.0$\pm$0.5    &  1.7$\pm$0.2	&  0.62$\pm$0.09 &Y \\
189 &18 46 46.57 &$-$2 35 42.9 &30.09 &$-$0.15 &   6.6$\pm$0.8   &   8.7$\pm$1.1   &   5.5$\pm$0.7    &  2.8$\pm$0.4	&  1.3$\pm$0.2 &Y \\
191 &18 46 48.73 &$-$2 40 17.9 &30.03 &$-$0.19 &   3.8$\pm$0.4   &  12$\pm$1   &   8.8$\pm$1.0    &  4.0$\pm$0.5	&  1.3$\pm$0.2 &N \\
192 &18 46 42.78 &$-$2 38 49.1 &30.04 &$-$0.16 &   0.95$\pm$0.16   &   9.0$\pm$1.1   &   6.6$\pm$0.8    &  2.9$\pm$0.4	&  0.79$\pm$0.13 &Y \\
193 &18 45 50.00 &$-$2 30 49.7 &30.06 &$+$0.10 &   0.54$\pm$0.10   &  10$\pm$2   &  13$\pm$3    &  9.8$\pm$2.0	&  4.7$\pm$1.0 &N \\
194 &18 45 45.93 &$-$2 36 36.7 &29.96 &$+$0.08 &    15$\pm$2     &  18$\pm$2   &  12$\pm$1    &  5.8$\pm$0.7	&  2.6$\pm$0.4 &Y \\
195 &18 46 31.55 &$-$2 32 34.5 &30.11 &$-$0.07 &   0.30$\pm$0.08   &   5.1$\pm$0.6   &   7.7$\pm$0.9    &  6.4$\pm$0.9	&  3.3$\pm$0.5 &Y \\
196 &18 45 47.56 &$-$2 37 22.7 &29.95 &$+$0.06 &   3.1$\pm$0.4   &  15$\pm$2   &  11$\pm$1    &  5.6$\pm$0.7	&  2.2$\pm$0.3 &Y \\
197 &18 46 52.33 &$-$2 40 02.5 &30.04 &$-$0.20 &   9.9$\pm$1.2   &  12$\pm$1   &   6.7$\pm$0.8    &  2.3$\pm$0.3	&  0.31$\pm$0.06 &Y \\
198 &18 45 50.16 &$-$2 35 01.3 &29.99 &$+$0.07 &   9.3$\pm$1.0   &  14$\pm$2   &   7.0$\pm$0.8    &  2.8$\pm$0.4	&  0.85$\pm$0.12 &Y \\
199 &18 46 48.81 &$-$2 41 17.3 &30.01 &$-$0.20 &  12$\pm$1   &  19$\pm$2   &  10$\pm$1    &  3.9$\pm$0.5	&  1.0$\pm$0.1 &Y \\
200 &18 46 10.48 &$-$2 46 23.5 &29.86 &$-$0.09 &   0.67$\pm$0.12   &  101$\pm$1   &   9.7$\pm$1.1    &  5.2$\pm$0.7	&  2.0$\pm$0.3 &N \\
201 &18 46 33.82 &$-$2 34 23.3 &30.09 &$-$0.09 &   5.4$\pm$0.6   &   8.9$\pm$1.0   &   5.1$\pm$0.6    &  2.1$\pm$0.3	&  0.77$\pm$0.11 &Y \\
202 &18 46 45.27 &$-$2 39 16.9 &30.04 &$-$0.17 &   2.3$\pm$0.4   &   7.2$\pm$1.1  &   4.4$\pm$0.7    &  1.7$\pm$0.3	&  0.46$\pm$0.09 &Y \\
203 &18 45 45.77 &$-$2 39 00.4 &29.93 &$+$0.05 &   1.5$\pm$0.2   &  12$\pm$1   &  12$\pm$1    &  7.1$\pm$0.9	&  3.0$\pm$0.4 &Y \\
205 &18 46 52.16 &$-$2 42 07.0 &30.01 &$-$0.22 &   6.9$\pm$0.8   &  13$\pm$1   &   8.5$\pm$0.9    &  3.7$\pm$0.5	&  1.2$\pm$0.2 &Y \\
207 &18 46 14.92 &$-$2 50 17.6 &29.81 &$-$0.14 &  13$\pm$2   &   9.2$\pm$1.3   &   4.6$\pm$0.6    &  1.8$\pm$0.3	&  0.61$\pm$0.10 &Y \\
208 &18 45 59.55 &$-$2 48 26.6 &29.81 &$-$0.07 &   0.89$\pm$0.16   &   6.6$\pm$0.8   &   4.7$\pm$0.5    &  1.8$\pm$0.3	&  0.42$\pm$0.08 &N \\
209 &18 45 42.44 &$-$2 31 26.2 &30.03 &$+$0.12 &   0.52$\pm$0.09   &  14$\pm$3   &  23$\pm$5    & 16$\pm$4	&  7.8$\pm$2.0 &N \\
210 &18 46 49.13 &$-$2 38 04.4 &30.06 &$-$0.17 &   9.5$\pm$1.1   &   9.3$\pm$1.0   &   5.0$\pm$0.6    &  2.0$\pm$0.3	&  0.47$\pm$0.07 &Y \\
212 &18 45 50.40 &$-$2 47 58.4 &29.80 &$-$0.03 &   1.3$\pm$0.2   &  10$\pm$1   &  12$\pm$1    &  8$\pm$1	&  3.6$\pm$0.5 &N \\
213 &18 46 51.47 &$-$2 38 02.6 &30.07 &$-$0.18 &   6.2$\pm$0.7   &  11$\pm$1   &   6.8$\pm$0.8    &  2.7$\pm$0.4	&  0.90$\pm$0.13 &Y \\
214 &18 45 43.98 &$-$2 45 11.7 &29.83 &$+$0.01 &   5.3$\pm$0.6   &  21$\pm$2   &  23$\pm$3    & 13$\pm$2	&  5.0$\pm$0.7 &N \\
215 &18 45 56.64 &$-$2 34 45.2 &30.01 &$+$0.05 &   1.1$\pm$0.1   &   7.5$\pm$0.9   &   9.0$\pm$1.0    &  5.8$\pm$0.8	&  2.7$\pm$0.4 &Y \\
217 &18 45 53.82 &$-$2 46 55.7 &29.82 &$-$0.04 &   0.79$\pm$0.14   &  13$\pm$1   &  10$\pm$1    &  4.8$\pm$0.6	&  1.9$\pm$0.3 &N \\
219 &18 46 01.96 &$-$2 30 47.6 &30.08 &$+$0.06 &   0.29$\pm$0.21   &   6.9$\pm$0.9   &   5.2$\pm$0.7    &  2.5$\pm$0.4	&  0.69$\pm$0.12 &Y \\
220 &18 46 14.12 &$-$2 32 10.8 &30.08 &$+$0.00 &   4.0$\pm$0.4   &  10$\pm$1   &   7.0$\pm$0.8    &  3.1$\pm$0.4    &  1.2$\pm$0.2 &Y \\
221 &18 46 10.14 &$-$2 51 21.1 &29.79 &$-$0.13 &   3.3$\pm$0.4   &  12$\pm$1   &   7.7$\pm$0.9    &  3.2$\pm$0.4    &  0.97$\pm$0.14 &Y \\
228 &18 45 50.23 &$-$2 48 25.2 &29.80 &$-$0.03 &   0.49$\pm$0.15   &  10$\pm$1   &   9.9$\pm$1.1    &  4.8$\pm$0.6	&  1.6$\pm$0.2 &Y \\
232 &18 45 35.61 &$-$2 39 04.1 &29.91 &$+$0.09 &   2.0$\pm$0.3   &  11$\pm$1   &   9.2$\pm$1.0    &  4.4$\pm$0.6	&  1.7$\pm$0.2 &N \\
242 &18 46 03.84 &$-$2 39 21.2 &29.96 &$-$0.02 & 7235$\pm$809 &1810$\pm$202  & 498$\pm$56   &348$\pm$45  &105$\pm$15 &Y  \\  
243 &18 45 59.45 &$-$2 45 05.8 &29.86 &$-$0.04 & 552$\pm$62  & 228$\pm$25  & 115$\pm$13   & 56$\pm$7    & 20$\pm$3 &Y \\ 
245 &18 46 11.25 &$-$2 41 56.2 &29.93 &$-$0.06 & 605$\pm$68  & 338$\pm$38  & 184$\pm$21   & 90$\pm$12   & 28$\pm$4 &Y \\ 
247 &18 46 17.08 &$-$2 36 43.5 &30.02 &$-$0.05 & 601$\pm$67  & 280$\pm$31  & 142$\pm$16   & 65$\pm$8    & 25$\pm$3 &Y \\ 
251 &18 45 54.67 &$-$2 42 53.2 &29.86 &$-$0.01 & 222$\pm$25  &  76$\pm$8   &  40$\pm$4    & 15$\pm$2    &  3.2$\pm$0.5 &Y \\ 
253 &18 46 01.75 &$-$2 45 27.7 &29.86 &$-$0.05 & 352$\pm$39  & 186$\pm$21  &  91$\pm$10   & 36$\pm$5	 & 13$\pm$2 &Y \\ 
254 &18 46 07.24 &$-$2 42 20.7 &29.92 &$-$0.05 &  99$\pm$14  & 173$\pm$24  & 116$\pm$16   & 49$\pm$7	 & 16$\pm$3 &Y \\
257 &18 46 06.94 &$-$2 42 58.6 &29.91 &$-$0.06 & 142$\pm$34  &  97$\pm$24  &  50$\pm$12   & 25$\pm$6    &  7.0$\pm$1.8 &Y \\
258 &18 45 55.72 &$-$2 42 31.1 &29.89 &$-$0.01 & 177$\pm$22  &  92$\pm$12  &  48$\pm$6    & 25$\pm$4	& 11$\pm$2 &Y \\ 
259 &18 46 07.95 &$-$2 43 23.8 &29.90 &$-$0.06 & 175$\pm$20  &  78$\pm$10  &  43$\pm$5    & 18$\pm$3	&  6.2$\pm$0.9 &Y \\
262 &18 46 04.86 &$-$2 42 44.1 &29.91 &$-$0.05 &  81$\pm$10  & 106$\pm$12  &  58$\pm$7    & 19$\pm$2    &  6.5$\pm$0.9 &Y \\
268 &18 45 45.64 &$-$2 31 52.3 &30.03 &$+$0.11 &  98$\pm$11  &  48$\pm$5   &  26$\pm$3    & 13$\pm$2    &  7.1$\pm$1.0 &Y \\
269 &18 45 44.56 &$-$2 32 18.4 &30.02 &$+$0.11 & 215$\pm$24  &  88$\pm$10  &  41$\pm$5    & 16$\pm$2	    &  4.5$\pm$0.6 &Y \\ 
270 &18 46 09.73 &$-$2 43 41.7 &29.90 &$-$0.07 & 116$\pm$14  &  60$\pm$7   &  26$\pm$3    & 10$\pm$1	&  2.6$\pm$0.4 &N \\
274 &18 46 08.27 &$-$2 48 04.0 &29.83 &$-$0.10 & 104$\pm$12  &  44$\pm$5   &  26$\pm$3    & 11$\pm$1	&  4.2$\pm$0.6 &N \\
275 &18 45 44.00 &$-$2 32 00.3 &30.03 &$+$0.11 & 123$\pm$14  &  61$\pm$7   &  30$\pm$3    & 13$\pm$22	&  4.1$\pm$0.6 &Y \\ 
276 &18 46 08.37 &$-$2 47 45.6 &29.84 &$-$0.10 &  19$\pm$2   &  16$\pm$2   &  13$\pm$2    &  9.0$\pm$1.2    &  3.9$\pm$0.6 &Y \\
278 &18 46 35.43 &$-$2 40 34.6 &30.00 &$-$0.14 &  33$\pm$4   &  36$\pm$4   &  26$\pm$3    & 15$\pm$2    &  6.5$\pm$0.9 &Y \\
280 &18 46 26.29 &$-$2 40 55.9 &29.98 &$-$0.11 &  62$\pm$7   &  46$\pm$5   &  23$\pm$3    &  8.9$\pm$1.2	    &  2.2$\pm$0.3 &Y \\
281 &18 46 01.29 &$-$2 46 23.4 &29.85 &$-$0.06 &  63$\pm$7   &  56$\pm$7   &  36$\pm$4    & 17$\pm$2	    &  5.0$\pm$0.7 &Y \\
282 &18 45 51.24 &$-$2 30 17.7 &30.01 &$+$0.10 & 120$\pm$13  &  50$\pm$6   &  24$\pm$3    &  9.5$\pm$1.2	&  3.6$\pm$0.5 &Y \\
285 &18 46 06.28 &$-$2 30 13.5 &30.10 &$+$0.04 &  24$\pm$3   &  19$\pm$2   &  11$\pm$1    &  5.3$\pm$0.7	&  2.2$\pm$0.3 &Y \\
286 &18 45 59.66 &$-$2 29 09.3 &30.10 &$+$0.08 &  21$\pm$2   &  32$\pm$4   &  26$\pm$3    & 14$\pm$2	&  5.8$\pm$0.8 &Y \\
288 &18 46 23.02 &$-$2 43 05.6 &29.94 &$-$0.12 &  46$\pm$5   &  33$\pm$4   &  17$\pm$2    &  6.8$\pm$0.9	&  2.4$\pm$0.3 &Y \\
289 &18 46 22.69 &$-$2 40 12.0 &29.98 &$-$0.09 &  34$\pm$4   &  34$\pm$4   &  19$\pm$2    &  8.2$\pm$1.1	&  2.7$\pm$0.4 &Y \\
291 &18 46 11.54 &$-$2 44 15.3 &29.90 &$-$0.08 &  26$\pm$3   & 9.0$\pm$1.5 & 4.0$\pm$0.8  &  1.4$\pm$0.4	&  0.14$\pm$0.12 &Y \\
292 &18 46 14.13 &$-$2 43 28.8 &29.91 &$-$0.09 &  10$\pm$2   & 9.4$\pm$1.9 & 4.3$\pm$0.9  &  1.9$\pm$0.4	&  0.39$\pm$0.12 &Y \\
293 &18 46 21.84 &$-$2 40 30.7 &29.97 &$-$0.09 &  35$\pm$4   &  25$\pm$3   &  13$\pm$1    &  4.5$\pm$0.6	&  0.95$\pm$0.17 &Y \\
294 &18 46 25.27 &$-$2 40 35.7 &29.98 &$-$0.11 &  21$\pm$2   &  18$\pm$2   &  12$\pm$1    &  5.9$\pm$0.8	&  2.6$\pm$0.4 &Y \\
301 &18 46 05.74 &$-$2 48 28.1 &29.82 &$-$0.09 &  30$\pm$3   &  28$\pm$3   &  22$\pm$2    & 12$\pm$2	        &  4.7$\pm$0.7 &N \\ 
305 &18 46 17.13 &$-$2 48 57.3 &29.84 &$-$0.14 &  24$\pm$3   &  17$\pm$2   & 9.3$\pm$1.0  &  4.3$\pm$0.6	&  1.7$\pm$0.2 &Y \\ 
306 &18 46 15.61 &$-$2 49 21.4 &29.83 &$-$0.14 &  12$\pm$1   &  12$\pm$1   & 6.8$\pm$0.8  &  2.6$\pm$0.3	&  0.70$\pm$0.10 &Y \\
307 &18 46 22.52 &$-$2 41 50.2 &29.95 &$-$0.10 &  10$\pm$2   & 8.3$\pm$1.8 & 5.1$\pm$1.1    &  2.3$\pm$0.5	&  0.79$\pm$0.19 &Y \\
309 &18 46 07.27 &$-$2 48 57.0 &29.82 &$-$0.10 &  16$\pm$2   & 8.2$\pm$0.9 & 4.5$\pm$0.6    &  1.9$\pm$0.3	&  1.0$\pm$0.1 &N \\
311 &18 46 03.84 &$-$2 36 31.1 &30.00 &$+$0.01 &  10$\pm$1   &  11$\pm$1   &  11$\pm$1    &  7.1$\pm$0.9	&  2.7$\pm$0.4 &Y \\
318 &18 46 05.61 &$-$2 35 17.5 &30.02 &$+$0.01 &  12$\pm$1   &  14$\pm$2   &  11$\pm$1    &  5.9$\pm$0.8	&  1.7$\pm$0.2 &Y \\
323 &18 46 03.43 &$-$2 35 20.1 &30.01 &$+$0.02 &  22$\pm$2   &  15$\pm$2   & 5.7$\pm$0.7    &  1.1$\pm$0.2	&  0.36$\pm$0.09 &Y \\
324 &18 45 57.37 &$-$2 36 50.2 &29.98 &$+$0.03 &  21$\pm$3   &  18$\pm$2   &   8.2$\pm$0.9    &  2.8$\pm$0.4	&  0.53$\pm$0.10 &Y \\
325 &18 45 56.12 &$-$2 48 38.9 &29.80 &$-$0.06 &  33$\pm$4   &  24$\pm$3   &  12$\pm$1    &  4.6$\pm$0.6	&  1.3$\pm$0.2 &Y \\
326 &18 46 27.13 &$-$2 39 48.3 &29.99 &$-$0.11 & 6.3$\pm$0.8   &   4.6$\pm$0.6   &   2.6$\pm$0.4    &  0.9$\pm$0.2        &  0.08$\pm$0.02 &Y \\
327 &18 46 19.36 &$-$2 45 06.4 &29.90 &$-$0.12 & 9.3$\pm$1.1   &   6.1$\pm$0.7   &   2.9$\pm$0.4    &  1.4$\pm$0.2        &  0.45$\pm$0.08 &Y \\
329 &18 46 05.85 &$-$2 30 33.3 &30.09 &$+$0.04 &  25$\pm$3   &  23$\pm$3   &  12$\pm$1	&  4.4$\pm$0.6      &  1.1$\pm$0.2 &Y \\ 
332 &18 46 21.60 &$-$2 50 08.4 &29.83 &$-$0.16 & 9.5$\pm$1.1   &   4.4$\pm$0.5   &   3.4$\pm$0.4    &  1.5$\pm$0.2        &  0.45$\pm$0.07 &Y \\
335 &18 45 56.04 &$-$2 37 47.5 &29.96 &$+$0.02 & 8.7$\pm$1.2   &   7.4$\pm$1.2   &   3.3$\pm$0.6    &  0.6$\pm$0.2        &  0.08$\pm$0.02 &Y \\
338 &18 46 49.17 &$-$2 36 09.5 &30.09 &$-$0.16 &  11$\pm$1   &  14$\pm$2   &   7.6$\pm$0.8    &  3.2$\pm$0.4        &  1.1$\pm$0.2 &Y \\
339 &18 46 20.46 &$-$2 46 34.1 &29.88 &$-$0.13 & 7.9$\pm$0.9   &   8.0$\pm$0.9   &   4.9$\pm$0.6    &  2.3$\pm$0.3        &  1.2$\pm$0.2 &Y \\
340 &18 46 40.22 &$-$2 38 11.5 &30.04 &$-$0.14 & 6.2$\pm$0.7   &   8.8$\pm$1.0   &   6.8$\pm$0.8    &  3.5$\pm$0.5        &  1.4$\pm$0.2 &Y \\
341 &18 46 40.16 &$-$2 35 34.1 &30.08 &$-$0.12 &  13$\pm$1   &   9.4$\pm$1.1   &   3.2$\pm$0.4	&  1.0$\pm$0.1    &  0.31$\pm$0.05 &Y \\
342 &18 45 42.77 &$-$2 37 22.4 &29.95 &$+$0.08 &  21$\pm$2   &  17$\pm$2   &   8.3$\pm$0.9    &  3.2$\pm$0.4        &  0.94$\pm$0.13 &Y \\
343 &18 46 15.14 &$-$2 51 14.3 &29.80 &$-$0.15 & 5.9$\pm$0.7   &   9.1$\pm$1.0   &   6.6$\pm$0.7    &  3.3$\pm$0.4        &  1.2$\pm$0.17 &Y \\
344 &18 46 25.04 &$-$2 48 46.4 &29.86 &$-$0.17 & 8.0$\pm$0.9   &   5.1$\pm$0.6   &   2.6$\pm$0.3    &  1.2$\pm$0.2        &  0.51$\pm$0.07 &Y \\
349 &18 46 49.98 &$-$2 42 49.3 &29.99 &$-$0.21 &  13$\pm$1   &   4.7$\pm$0.5   &   2.0$\pm$0.3	&  0.96$\pm$0.15    &  0.35$\pm$0.06 &Y \\
352 &18 46 13.30 &$-$2 32 35.8 &30.07 &$+$0.00 & 2.4$\pm$0.3   &   4.4$\pm$0.5   &   3.3$\pm$0.4    &  1.6$\pm$0.2        &  0.46$\pm$0.06 &Y \\
357 &18 45 58.60 &$-$2 35 02.0 &30.01 &$+$0.04 &0.31$\pm$0.09   &   2.8$\pm$0.7   &   5.9$\pm$1.3    &  5.5$\pm$1.2      &  2.9$\pm$0.7 &Y \\
\hline	   		
\label{tflux}		        				    		 								
\end{longtable}
\end{scriptsize}		        				    		 
\renewcommand{\thefootnote}{\arabic{footnote}}

\clearpage

\begin{table*}
\caption[]{Mid-infrared fluxes used for the SED fitting} 
\label{tmir}
\begin{tabular}{rccc}
\hline
&&
\multicolumn{1}{c}{$\lambda$} &
\multicolumn{1}{c}{$S$}  
\\
\multicolumn{1}{c}{\# Id.$^a$} &
\multicolumn{1}{c}{Instrument} &
\multicolumn{1}{c}{($\mu$m)} &
\multicolumn{1}{c}{(Jy)}  
\\  
\hline
132 & MIPSGAL     &  24  & 0.44 \\
135 & MIPSGAL     &  24  & 0.27 \\
143 & MIPSGAL     &  24  & 0.024 \\
147 & MIPSGAL     &  24  & 0.047 \\
242 & MSX     &  21  & 1340 \\
243 & MSX     &  21  & 19 \\
245 & MSX     &  21  & 26 \\
247 & WISE    &  22  & 16   \\
251 & MIPSGAL     &  24  & 4.6 \\
253 & WISE    &  22  & 9.9   \\
257 & WISE    &  22  & 4.3   \\
258 & MIPSGAL     &  24  & 1.5 \\
259 & MIPSGAL     &  24  & 1.2 \\
268 & MIPSGAL     &  24  & 2.0 \\
269 & MIPSGAL     &  24  & 3.9 \\
270 & WISE    &  22  & 1.3   \\
274 & WISE    &  22  & 2.7   \\
275 & MIPSGAL     &  24  & 1.02 \\
282 & MIPSGAL     &  24  & 1.1 \\
285 & MIPSGAL     &  24  & 0.74 \\
288 & MIPSGAL     &  24  & 0.52 \\
291 & MIPSGAL     &  24  & 0.021 \\
305 & MIPSGAL     &  24  & 0.31 \\
309 & WISE    &  22  & 0.49   \\
327 & MIPSGAL     &  24  & 0.20 \\
342 & MIPSGAL     &  24  & 0.16 \\
344 & MSX     &  21  & 8.0 \\
349 & MIPSGAL     &  24  & 0.39 \\
\hline
\end{tabular} 
\\
$^a$ The number corresponds to the \hi\ identification number
(Table~1). 
\end{table*}

\clearpage

\onecolumn
\begin{scriptsize}
\label{tsed}
\begin{center}
\begin{longtable}{lcccccc}
\caption[] {Properties of the \hi\ sources in the G29.96$-$0.02 cloud from the
SED fitting} \\
\hline 
&&\multicolumn{1}{c}{$\theta$} &
\multicolumn{1}{c}{$T$} &
\multicolumn{1}{c}{$M_{\rm env}$} &
\multicolumn{1}{c}{$\Sigma$} &  
\multicolumn{1}{c}{$L_{\rm bol}$} 
\\
\multicolumn{1}{c}{\# Id.} &
\multicolumn{1}{c}{$\beta$}&
\multicolumn{1}{c}{($\arcsec$)} &
\multicolumn{1}{c}{(K)} &
\multicolumn{1}{c}{($M_\odot$)} &
\multicolumn{1}{c}{(g\,cm$^{-2}$)} &
\multicolumn{1}{c}{($L_\odot$)}\\
\hline \\[-2ex]
\endfirsthead
\multicolumn{7}{c}{{\tablename} \thetable{} -- Continued} \\[0.5ex]
\hline \\[-2ex]

&&\multicolumn{1}{c}{$\theta$} &
\multicolumn{1}{c}{$T$} &
\multicolumn{1}{c}{$M_{\rm env}$} &
\multicolumn{1}{c}{$\Sigma$} &  
\multicolumn{1}{c}{$L_{\rm bol}$} 
\\
\multicolumn{1}{c}{\# Id.} &
\multicolumn{1}{c}{$\beta$}&
\multicolumn{1}{c}{($\arcsec$)} 
&\multicolumn{1}{c}{(K)} &
\multicolumn{1}{c}{($M_\odot$)} &
\multicolumn{1}{c}{(g\,cm$^{-2}$)} &
\multicolumn{1}{c}{($L_\odot$)}\\
\hline \\[-2ex]
\endhead
  1 &  0.8  &19.44 &  41.9  & 645 & 0.50 &  3823   \\
  2 &  1.4  &27.48 &  27.4  & 966 & 0.38 &  7577   \\
  4 &  1.2  &21.81 &  15.8  &2426 & 1.5  &   705   \\
  5 &  2.2  &18.75 &  15.8  & 216 & 0.18 &   438   \\
  6 &  1.8  &18.30 &  15.8  & 457 & 0.40 &   427   \\
  7 &  2.2  &17.67 &  15.8  & 172 & 0.16 &   363   \\
  8 &  1.4  &18.91 &  18.7  & 257 & 0.21 &   272   \\
  9 &  2.0  &18.19 &  12.9  & 272 & 0.24 &   111   \\
 11 &  1.8  &23.06 &  21.6  &  41 & 0.02 &   233   \\
 12 &  2.6  &22.26 &  10.0  & 513 & 0.31 &   123   \\
 13 &  1.0  &25.56 &  18.7  & 543 & 0.25 &   253   \\
 14 &  0.8  &26.24 &  27.4  & 204 & 0.09 &   400   \\
 16 &  0.6  &23.91 &  24.5  & 323 & 0.17 &   188   \\
 17 &  0.4  &27.48 &  27.4  & 724 & 0.28 &   405   \\
 18 &  1.2  &17.28 &  15.8  & 257 & 0.25 &    75   \\
 19 &  1.6  &24.07 &  18.7  & 115 & 0.06 &   185   \\
 20 &  1.8  &26.33 &  12.9  & 363 & 0.15 &   103   \\
 21 &  1.8  &27.48 &  18.7  &  91 & 0.04 &   226   \\
 22 &  1.6  &26.38 &  21.6  &  68 & 0.03 &   247   \\
 24 &  1.8  &24.39 &  18.7  &  72 & 0.04 &   180   \\
 25 &  1.4  &26.15 &  21.6  & 172 & 0.07 &   197   \\
 26 &  2.0  &24.27 &  12.9  & 257 & 0.13 &   105   \\
 27 &  1.2  &23.59 &  24.5  & 122 & 0.06 &   232   \\
 31 &  2.6  &21.48 &  10.0  & 431 & 0.28 &   104   \\
 33 &  1.8  &23.80 &  21.6  &  48 & 0.03 &   277   \\
 34 &  1.6  &26.87 &  18.7  & 108 & 0.04 &   175   \\
 35 &  1.8  &24.00 &  24.5  &  36 & 0.02 &   243   \\
 37 &  1.2  &19.97 &  21.6  & 129 & 0.10 &   190   \\
 38 &  1.4  &23.16 &  24.5  &  46 & 0.03 &   208   \\
 40 &  1.2  &18.57 &  27.4  &  54 & 0.05 &   222   \\
 42 &  1.4  &26.38 &  24.5  &  46 & 0.02 &   208   \\
 44 &  0.4  &25.37 &  33.2  & 172 & 0.08 &   315   \\
 45 &  1.2  &23.28 &  21.6  & 243 & 0.13 &   193   \\
 48 &  1.4  &22.49 &  27.4  &  32 & 0.02 &   269   \\
 49 &  1.0  &20.18 &  27.4  &  77 & 0.06 &   201   \\
 54 &  2.4  &22.40 &  21.6  & 7.7 & 0.005&   183   \\
 55 &  2.4  &21.59 &  12.9  & 108 & 0.07 &    96   \\
 61 &  2.4  &27.20 &  15.8  &  54 & 0.02 &   176   \\
 62 &  0.6  &23.19 &  27.4  & 122 & 0.07 &   136   \\
 64 &  1.0  &19.60 &  24.5  & 122 & 0.09 &   170   \\
 65 &  1.4  &24.80 &  24.5  &  54 & 0.03 &   247   \\
 66 &  0.8  &24.30 &  27.4  &  91 & 0.05 &   158   \\
 69 &  1.2  &25.05 &  30.3  &  58 & 0.03 &   494   \\
 70 &  2.4  &24.43 &  10.0  & 645 & 0.32 &   111   \\
 73 &  2.6  &24.41 &  10.0  & 305 & 0.15 &    70   \\
 74 &  2.2  &24.87 &  15.8  &  72 & 0.03 &   152   \\
 75 &  2.6  &23.23 &  15.8  &  23 & 0.01 &   114   \\
 77 &  1.0  &21.61 &  27.4  &  58 & 0.04 &   181   \\
 79 &  2.0  &27.48 &  24.5  &  32 & 0.01 &   287   \\
 81 &  0.8  &21.63 &  30.3  &  77 & 0.05 &   244   \\
 83 &  2.4  &23.39 &  12.9  &  77 & 0.04 &    67   \\
 86 &  2.2  &12.81 &  15.8  &  27 & 0.05 &    58   \\
 87 &  1.6  &25.50 &  21.6  &  38 & 0.02 &   139   \\
 88 &  2.0  &27.48 &  18.7  &  48 & 0.02 &   187   \\
 90 &  1.2  &27.11 &  24.5  &  61 & 0.02 &   173   \\
 98 &  0.8  &27.48 &  24.5  & 513 & 0.20 &   336   \\
 99 &  0.4  &22.96 &  33.2  & 108 & 0.06 &   199   \\
100 &  2.4  &16.04 &  24.5  & 3.8 & 0.004 &  206   \\
104 &  2.0  &25.49 &  18.7  &  41 & 0.02 &   157   \\
109 &  1.6  &19.07 &  21.6  &  43 & 0.04 &   156   \\
111 &  1.2  &21.09 &  21.6  &  86 & 0.06 &   127   \\
113 &  2.0  &19.03 &  12.9  & 122 & 0.10 &    50   \\
122 &  1.8  &26.27 &  27.4  & 575 & 0.25 & 12985    \\
123 &  1.4  &24.86 &  21.6  &3234 & 1.54 &  7446   \\
124 &  2.6  &13.86 &  39.0  & 102 & 0.16 &  3544   \\
125 &  0.6  &23.61 &  41.9  &1288 & 0.68 & 11232   \\
126 &  0.6  &19.55 &  27.4  &3844 & 3.0  &  4647    \\
127 &  1.4  &28.26 &  30.3  & 645 & 0.24 &  8845    \\
129 &  0.4  &21.03 &  30.3  &2161 & 1.4 &   2651    \\
130 &  2.2  &26.26 &  12.9  &2883 & 1.2 &   1739   \\
131 &  1.4  &20.69 &  15.8  &2041 & 1.4 &    821   \\
132 &  1.2  &24.30 &  41.9  &  81 & 0.04 &  3765   \\
133 &  1.8  &19.34 &  15.8  & 767 & 0.60 &   677   \\
134 &  1.2  &21.90 &  15.8  &2883 & 1.8 &   798   \\
135 &  2.0  &18.32 &  39.0  & 4.3 & 0.004 &  1369   \\
136 &  1.8  &21.60 &  15.8  & 363 & 0.23 &   339   \\
137 &  1.8  &28.28 &  21.6  & 431 & 0.16 &  2338   \\
138 &  2.2  &21.19 &  21.6  &  48 & 0.03 &   713   \\
139 &  1.4  &24.27 &  36.1  &  86 & 0.04 &  1555   \\
141 &  2.4  &12.37 &  24.5  & 6.5 & 0.01 &   345   \\
142 &  2.0  &27.48 &  15.8  & 343 & 0.13 &   461   \\
143 &  1.0  &14.80 &  44.8  & 3.8 & 0.005 &  140   \\
144 &  0.6  &26.62 &  39.0  & 115 & 0.05 &   720   \\
147 &  1.6  &24.34 &  36.1  &  16 & 0.008 &  1041  \\
148 &  1.4  &26.16 &  18.7  & 767 & 0.33 &   811   \\
149 &  1.8  &22.43 &  18.7  & 384 & 0.23 &   919   \\
150 &  2.2  &25.70 &  15.8  & 431 & 0.19 &   915   \\
151 &  1.6  &28.24 &  21.6  & 609 & 0.23 &   312   \\
152 &  2.2  &30.00 &  18.7  & 162 & 0.05 &   930   \\
153 &  1.6  &30.00 &  36.1  &  51 & 0.02 &  1528   \\
155 &  0.4  &24.92 &  24.5  &1147 & 0.54 &   337   \\
159 &  1.4  &26.74 &  18.7  & 861 & 0.35 &   885   \\
160 &  2.6  &25.77 &  30.3  &  12 & 0.005 &  894   \\
161 &  2.6  &23.14 &  21.6  &  32 & 0.02 &   237   \\
162 &  1.6  &24.09 &  21.6  & 243 & 0.12 &   258   \\
163 &  2.0  &16.97 &  21.6  & 136 & 0.14 &   265   \\
164 &  0.6  &17.39 &  21.6  & 484 & 0.47 &   200   \\
165 &  0.8  &24.47 &  30.3  & 272 & 0.13 &   594   \\
167 &  1.6  &23.25 &  27.4  &  51 & 0.03 &   702   \\
168 &  1.4  &30.00 &  21.6  & 288 & 0.09 &   661   \\
169 &  0.4  &17.80 &  33.2  & 243 & 0.23 &   445   \\
170 &  1.4  &25.20 &  27.4  & 153 & 0.07 &   563   \\
171 &  1.6  &21.54 &  15.8  & 645 & 0.41 &   405   \\
172 &  0.4  &25.67 &  27.4  &5432 & 2.4  &   860   \\
175 &  0.4  &17.39 &  21.6  &2426 & 2.4  &   279   \\
177 &  1.0  &30.00 &  27.4  & 323 & 0.11 &  1017   \\
178 &  1.4  &25.93 &  21.6  & 243 & 0.11 &   559   \\
179 &  1.0  &22.50 &  36.1  &  38 & 0.02 &   425   \\
180 &  1.4  &22.70 &  30.3  &  72 & 0.04 &   617   \\
182 &  2.2  &29.25 &  15.8  & 323 & 0.11 &   686   \\
184 &  0.4  &28.02 &  24.5  &2426 & 0.91 &   469   \\
185 &  1.8  &24.95 &  24.5  &  31 & 0.01 &   363   \\
186 &  1.8  &27.68 &  27.4  &  27 & 0.01 &   362   \\
187 &  1.8  &30.00 &  27.4  &  97 & 0.03 &   639   \\
188 &  1.4  &24.14 &  24.5  &  48 & 0.02 &   220   \\
189 &  0.6  &24.57 &  36.1  & 108 & 0.05 &   476   \\
191 &  1.8  &25.75 &  21.6  &  81 & 0.04 &   466   \\
192 &  2.0  &24.94 &  18.7  &  61 & 0.03 &   234   \\
193 &  2.2  &16.71 &  21.6  & 288 & 0.30 &   314   \\
194 &  0.6  &28.03 &  39.0  & 216 & 0.08 &  1094   \\
195 &  0.6  &18.93 &  18.7  & 861 & 0.71 &   184   \\
196 &  1.4  &30.00 &  21.6  & 204 & 0.07 &   470   \\
197 &  2.6  &27.99 &  30.3  &   6 & 0.002 &  724   \\
198 &  2.0  &28.68 &  24.5  &  34 & 0.01 &   668   \\
199 &  2.0  &29.81 &  24.5  &  46 & 0.02 &   891   \\
200 &  2.2  &26.78 &  21.6  &  91 & 0.04 &   260   \\
201 &  1.4  &30.00 &  30.3  &  41 & 0.01 &   422   \\
202 &  2.0  &24.03 &  21.6  &  27 & 0.01 &   249   \\
203 &  1.0  &25.12 &  21.6  & 384 & 0.18 &   368   \\
205 &  1.4  &30.00 &  30.3  &  68 & 0.02 &   596   \\
207 &  1.4  &25.40 &  36.1  &  24 & 0.01 &   694   \\
208 &  2.2  &22.35 &  18.7  &  29 & 0.02 &   173   \\
209 &  0.4  &23.35 &  18.7  &2883 & 1.6  &   423   \\
210 &  2.0  &29.79 &  27.4  &  16 & 0.005&   618    \\
212 &  0.8  &17.82 &  21.6  & 543 & 0.50 &   340    \\
213 &  1.8  &29.56 &  24.5  &  43 & 0.02 &   513    \\
214 &  2.6  &29.64 &  12.9  & 407 & 0.14 &   456    \\
215 &  0.8  &25.20 &  21.6  & 431 & 0.20 &   270    \\
217 &  2.4  &30.00 &  15.8  & 102 & 0.03 &   331    \\
219 &  2.4  &28.04 &  15.8  &  48 & 0.02 &   154    \\
220 &  1.4  &30.00 &  24.5  &  86 & 0.03 &   391    \\
221 &  2.0  &28.99 &  21.6  &  48 & 0.02 &   441    \\
228 &  2.2  &27.80 &  15.8  & 129 & 0.05 &   267   \\
232 &  1.6  &30.00 &  24.5  & 108 & 0.04 &   344   \\
242 &  0.8  &19.39 &  76.7  &2883 & 2.3  &791095   \\
243 &  1.4  &14.84 &  70.9  & 431 & 0.58 & 39363   \\
245 &  2.6  &23.75 &  73.8  & 229 & 0.12 & 47243   \\
247 &  0.6  &23.65 &  56.4  &1215 & 0.64 & 41592   \\
251 &  1.6  &21.38 &  56.4  &  72 & 0.05 & 13062   \\
253 &  1.8  &27.42 &  62.2  & 193 & 0.08 & 23156   \\
254 &  1.4  &22.55 &  27.4  &1023 & 0.59 &  7999   \\
257 &  0.6  &19.57 &  59.3  & 457 & 0.35 & 11617   \\
258 &  0.8  &21.06 &  44.8  & 513 & 0.34 & 10648   \\
259 &  1.2  &23.91 &  41.9  & 229 & 0.12 & 10559   \\
262 &  2.4  &22.82 &  44.8  &  77 & 0.04 &  6097   \\
268 &  0.4  &16.48 &  53.5  & 384 & 0.42 &  5751   \\
269 &  1.4  &26.91 &  44.8  & 129 & 0.05 & 15111   \\
270 &  1.4  &25.58 &  53.5  &  72 & 0.03 &  8826   \\
274 &  0.6  &25.40 &  56.4  & 216 & 0.10 &  7396   \\
275 &  1.2  &29.45 &  41.9  & 162 & 0.06 &  7511   \\
276 &  0.4  &11.75 &  39.0  & 323 & 0.69 &  1204    \\
278 &  0.4  &13.71 &  39.0  & 609 & 0.96 &  2268    \\
280 &  1.8  &22.85 &  41.9  &  54 & 0.03 &  3690    \\
281 &  1.4  &27.87 &  41.9  & 204 & 0.08 &  4078    \\
282 &  1.0  &27.69 &  44.8  & 162 & 0.06 &  5954    \\
285 &  0.4  &15.01 &  53.5  & 144 & 0.19 &  1966    \\
286 &  0.6  &14.34 &  36.1  & 645 & 0.92 &  1637    \\
288 &  0.8  &23.29 &  47.7  & 144 & 0.08 &  3018    \\
289 &  1.2  &26.84 &  39.0  & 136 & 0.06 &  2195    \\
291 &  2.6  &24.19 &  30.3  & 3.8 & 0.002&  1390   \\
292 &  2.0  &18.77 &  27.4  &  15 & 0.01 &   584   \\
293 &  2.2  &27.86 &  27.4  &  31 & 0.01 &  1965   \\
294 &  0.6  &23.29 &  41.9  & 193 & 0.11 &  1375   \\
301 &  0.4  &30.00 &  41.9  & 513 & 0.17 &  1957   \\
305 &  0.6  &30.00 &  47.7  & 102 & 0.03 &  1619   \\
306 &  1.6  &23.90 &  30.3  &  32 & 0.02 &   779   \\
307 &  1.0  &21.63 &  36.1  &  48 & 0.03 &   604    \\
309 &  0.4  &24.86 &  59.3  &  51 & 0.02 &  1206    \\
311 &  0.4  &13.35 &  36.1  & 323 & 0.53 &   717    \\
318 &  0.8  &15.52 &  36.1  & 153 & 0.19 &   875   \\
323 &  2.6  &27.19 &  27.4  & 6.5 & 0.003&  1195   \\
324 &  2.6  &24.65 &  24.5  &  14 & 0.007&  1224   \\
325 &  1.6  &27.05 &  39.0  &  41 & 0.02 &  1978   \\
326 &  2.6  &24.05 &  24.5  & 3.8 & 0.002&   344   \\
327 &  0.6  &22.58 &  50.6  &  31 & 0.02 &   634   \\
329 &  2.0  &26.47 &  27.4  &  41 & 0.02 &  1554   \\
332 &  0.8  &26.97 &  44.8  &  27 & 0.01 &   565   \\
335 &  2.4  &26.01 &  27.4  & 3.8 & 0.002&   421   \\
338 &  1.4  &30.00 &  33.2  &  51 & 0.02 &   778   \\
339 &  0.6  &25.09 &  39.0  &  81 & 0.04 &   509   \\
340 &  0.8  &21.85 &  33.2  & 115 & 0.07 &   471   \\
341 &  2.2  &30.00 &  33.2  & 5.8 & 0.002&   688   \\
342 &  1.0  &25.91 &  41.9  &  58 & 0.03 &  1512   \\
343 &  1.0  &26.35 &  30.3  &  91 & 0.04 &   474   \\
344 &  0.8  &25.27 &  44.8  &  26 & 0.01 &   505   \\
349 &  1.0  &30.00 &  50.6  &  12 & 0.004&   821    \\
352 &  1.4  &17.62 &  27.4  &  32 & 0.03 &   213    \\
357 &  0.4  & 9.66 &  21.6  & 683 & 2.2  &   140    \\
\hline  	    						     
\end{longtable} 						     
\end{center}							     
\end{scriptsize}

\end{document}